\documentclass[a4paper,fleqn,usenatbib]{mnras}

\usepackage{newtxtext,newtxmath}
\usepackage[T1]{fontenc}
\usepackage{ae,aecompl}

\usepackage{graphicx}	% Including figure files
\usepackage{amsmath}	% Advanced maths commands
\usepackage{amssymb}	% Extra maths symbols

\title[Radiation pressure on dust at high redshift]{Radiation-pressure-driven
dust transport to galaxy halos at $z\sim 10$}

\author[H. Hirashita and A. K. Inoue]{
Hiroyuki Hirashita$^1$\thanks{E-mail: hirashita@asiaa.sinica.edu.tw} and
Akio K. Inoue$^{2,3,4}$
\\
$^{1}$Institute of Astronomy and Astrophysics, Academia Sinica,
Astronomy-Mathematics Building, AS/NTU,
No.\ 1, Sec.\ 4, Roosevelt Road, Taipei 10617, Taiwan \\
$^{2}$Department of Environmental Science and Technology,
Faculty of Design Technology, Osaka Sangyo University, 3-1-1,
Nakagaito, Daito, Osaka 574-8530, Japan\\
$^3$Department of Physics, School of Advanced Science and Engineering,
Waseda University, 3-4-1, Okubo, Shinjuku, Tokyo 169-8555, Japan\\
$^4$ Waseda Research Institute for Science and Engineering, 3-4-1,
Okubo, Shinjuku, Tokyo 169-8555, Japan
}

\date{Accepted XXX. Received YYY; in original form ZZZ}

\pubyear{2019}

\begin{document}
\label{firstpage}
\pagerange{\pageref{firstpage}--\pageref{lastpage}}
\maketitle

\begin{abstract}
The origin of dust in galaxy halos or in the circum-galactic medium (CGM)
is still a mystery. We investigate if the radiation pressure in high-redshift
($z\sim 10$)
galaxies can efficiently transport dust to halos. To clarify the
first dust enrichment of galaxy halos in the early Universe,
we solve the motion of a dust grain
considering radiation pressure, gas drag, and gravity
in the vertical direction of the galactic disc. Radiation pressure is estimated
in a consistent manner with the stellar spectra and dust extinction.
As a consequence, we find that dust grains with radii $a\sim 0.1~\micron$
successfully escape from the galactic disc if the ongoing star formation
episode converts more than 15 per cent of
the baryon content into stars and lasts $\gtrsim 30$ Myr, while larger
and smaller grains are trapped in the disc because of gravity and gas drag,
respectively.
We also show that grain charge significantly enhances
gas drag at a few--10 scale heights of the galactic disc, where the
grain velocities are suppressed to $\sim 1$ km s$^{-1}$.
There is an optimum dust-to-gas ratio ($\sim 10^{-3}$) in the galactic disc
and an optimum virial mass $\sim 10^{10}$--$10^{11}$ M$_{\sun}$ for the
transport of $a\sim 0.1~\micron$ grains to the halo. We conclude that
early dust enrichment of galaxy halos at $z\gtrsim 10$ is important for the
origin of dust in the CGM.
\end{abstract}


%%%%%%%%%%%%%%%%% BODY OF PAPER %%%%%%%%%%%%%%%%%%
\section{Introduction}
\label{sec:intro}

Dust is known to exist in a wide volume in the Universe,
not only in the interstellar medium (ISM) but also in the
circum-galactic and intergalactic medium (CGM and IGM)
\citep[e.g.][]{Menard:2010aa}.
The dust content in galaxy halos (or the CGM; hereafter, we simply
use words `halo' to indicate the circum-galactic environment)
gives us a clue to the transport of dust from the ISM to the IGM, since
halos are the interface between the ISM and the IGM.
\citet[][]{Menard:2010aa} detected
reddening
in galaxy halos using the cross-correlation between
the galaxy position and the reddening of background quasi-stellar objects
(QSOs) for
a large sample of galaxies taken by the Sloan Digital Sky Survey
(SDSS; \citealt{York:2000aa}). The median redshift of their
sample is $z\sim 0.3$ ($z$ denotes the redshift).
They detected reddening up to
a radius of several Mpc from the galaxy centre.
\citet{Peek:2015aa} applied basically the same method to nearby
galaxies ($z\sim 0.05$), and found a similar radial profile of
reddening to the one found in \citet{Menard:2010aa}.
\citet{Masaki:2012aa} confirmed the observationally suggested
large extent of dust distribution in galaxy halos by comparing their
analytic halo model with \citet{Menard:2010aa}'s data.

Since dust in the CGM and IGM affects the opacity toward distant
objects in the Universe
\citep{Aguirre:1999aa}, it is important to clarify the origin and evolution
of dust in the cosmic volume.
Dust in galaxy halos is also of fundamental importance in the total dust budget
in galaxies and in the Universe, since \citet{Menard:2010aa} estimate that
the dust mass in a galaxy halo is on average
comparable to that in a galaxy disc.
Moreover, as \citet{Inoue:2003ab,Inoue:2004aa,Inoue:2010aa} argued,
dust could affect the thermal state of the IGM through
photoelectric heating. They constrained the grain size and
the dust-to-gas ratio in the IGM using the observed thermal
history of the IGM, although those two quantities are
degenerate in such a way that small grains require
smaller abundance of dust.

Dust forms and evolves mainly in the
ISM through various processes \citep{Asano:2013aa}.
Since a galaxy is
not a closed system, the ISM interacts with the CGM and IGM
through outflow
driven by supernovae (SNe) and active galactic nuclei (AGNs)
\citep[e.g.,][]{Veilleux:2005aa},
and through inflow driven by cooling and/or gravity
\citep[e.g.,][]{Keres:2005aa}. The outflow
could also transport the interstellar dust to the CGM and IGM
\citep{Zu:2011aa,McKinnon:2016aa,Hou:2017aa,Aoyama:2018aa}.
 The exchange of materials between the ISM
and the CGM/IGM, thus, affects the dust abundances in both the
ISM and the CGM/IGM.

Dust motion from the ISM to the CGM/IGM could also be
induced by a non-hydrodynamical process.
Radiation pressure
from stars in a galaxy could also drive the interstellar
dust outward, supplying the dust to the CGM and IGM.
In spite of some theoretical calculations of this process
\citep{Ferrara:1991aa,Bianchi:2005aa}, it is generally difficult to
directly simulate the dust motion on galactic (or larger) scales, since
the decoupling between dust and gas (i.e.\ the multi-fluid nature)
is essential.
\citet{Bekki:2015aa}, using multi-component particle methods,
have shown that the radial dust density profile is affected by
radiation pressure, but they did not focus on the
ejection of dust particles from the galactic disc to the halo.
Therefore, the current understanding of the CGM and IGM dust abundances
are primarily based on the hydrodynamical effects (i.e.\ dust motion
driven by hydrodynamical outflows). Further studies are necessary
to investigate dust transport by radiation pressure.

\citet{Barsella:1989aa} considered radiation pressure, gas drag (friction), and
gravity and calculated the force field around
a galaxy. They found that graphite grains have an outward force field due to
efficient radiation pressure; thus, they expected that graphite grains are
expelled out of the galaxy.
\citet{Davies:1998aa} also considered the motion in the gravitational
potential typical of a disc galaxy composed of stars, gas, dust and
dark halo. {They showed that dust grains with $a\sim 0.1~\micron$
($a$ is the grain radius)
can be ejected from the galactic disc but that the dust motion is sensitive to the
disc opacity. Small grains with $a<0.01~\micron$ stay relatively near to
the disc because they are inefficient in receiving radiation pressure
(i.e.\ they have small absorbing and scattering efficiencies).}
\citet{Simonsen:1999aa} showed that the velocities of escaping grains
from a galactic disc are not strongly dependent on the grain size. They also
mentioned that small grains could be slowed down by gas drag
if the gas density in the halo is as high as expected for high-redshift
galaxies.
\citet{Ferrara:1991aa} investigated the motion of dust pushed by
radiation pressure toward the halo in the gravitational potential
and the gas density profile typical of nearby spiral galaxies.
They showed that dust grains obtain velocities in excess of
100 km s$^{-1}$ \citep[see also][]{Shustov:1995aa}, and that
grains with $a\sim 0.1~\micron$
survive against sputtering in the hot halo.
\cite{Bianchi:2005aa}, using the IGM density distribution
at $z\sim 3$ in a cosmological simulation, showed that dust grains
transported to the IGM contribute to the metal enrichment there through
sputtering. They argued that large ($a\gtrsim 0.1~\micron$) grains
are preferentially injected in the low-density IGM, since smaller
grains are decelerated by gas drag in denser regions near galaxies.

The above studies focus on relatively low redshifts.
In considering the origin of dust in the IGM, systematic studies
starting from high redshift are crucial. Moreover, recent sensitive
submillimetre and millimetre observations by ALMA have found some dusty `normal'
galaxies at $z>7$ \citep{Watson:2015aa,Laporte:2017aa,Hashimoto:2018ab,Tamura:2019aa}.
These galaxies could be the first sources of the dust in the IGM and CGM.
This means that the first enrichment of dust on a wide
scale in the Universe could have occurred at $z\gtrsim 7$.
The above previous studies did not target such
high redshift galaxies. Thus, it is worth investigating dust ejection from
galaxies at $z\gtrsim 7$ as the `first' dust sources in
the CGM and IGM.

We expect that high-redshift galaxies manifest some differences
in radiation-pressure-driven
dust motion from low-redshift ones. A higher matter density at high redshift
would cause stronger gravity.
Since gravity could counteract radiation pressure, dust grains, especially large ones,
may not efficiently escape out of the galactic disc.
The higher baryon density potentially causes stronger gas drag; thus,
the grain velocities may become slower or the grains could even be trapped
in the galactic disc.
On the other hand, if star formation occurs in a compact
region, the stellar surface brightness (intensity) becomes higher. This
makes radiation pressure stronger.
The lower metal content in high-redshift galaxies could predict
stronger radiation pressure because less fraction of stellar
light is extinguished. Denser gas density, on the contrary, will
raise the extinction optical depth.

The goal of this paper is to examine the above effects regarding
dust motion systematically for galaxies at redshifts as high as $\gtrsim 7$.
We put particular focus on whether or not the dust grains in the
galactic disc (or the ISM) are successfully pushed and transported to the
galaxy halo (or CGM) by radiation pressure in spite of
the counteracting forces (gas drag and gravity).
This transport is referred to as \textit{grain escape}.
As mentioned above, it is not easy to solve dust dynamics
in a cosmological simulation, although
there are some codes that are capable of treating gas--grain decoupling
\citep{Bekki:2015aa,McKinnon:2018aa}.
Thus, some analytic approaches that focus on high-redshift
galaxies would be useful. In this paper, we analytically model the three
relevant forces -- radiation pressure, gas drag, and gravity --
consistently, focusing
on relevant redshift scalings. In this way, we will be able to
address the importance of radiation pressure on the dust motion
in a physical condition typical of high-redshift galaxies,
although we apply some simplifications to make the problem
analytically tractable.
This paper also provides a useful step toward
future direct simulations of dust motion or a viable tool to be used
in semi-analytic galaxy evolution models.
We also neglect small-scale structures in galactic disc; thus,
this work is
complementary to the studies that focus on radiation feedback
on small scales such as H\,\textsc{ii} regions
\citep[e.g.][]{Akimkin:2015aa,Akimkin:2017aa,Ishiki:2018aa}.

The paper is organized as follows. We formulate the model
in Section~\ref{sec:model}.
We show the results in Section~\ref{sec:result}. We further discuss
the model predictions and their implications in
Section \ref{sec:discussion}.
Finally we conclude in Section \ref{sec:conclusion}.
Throughout this paper, we adopt $H_0=70$ km s$^{-1}$ Mpc$^{-1}$
(the Hubble constant at $z=0$),
$\Omega_\mathrm{M}=0.3$ (the matter density normalized to
the critical density),
$\Omega_\Lambda =0.7$ (the cosmological constant),
and $\Omega_\mathrm{b}=0.04$ (the baryon density normalized to the critical
density) for the
cosmological parameters.

\section{Model}\label{sec:model}

We construct a framework that describes the dust motion in a high-redshift
galaxy with an explicit dependence on the redshift.
The entire formulation is composed of
a galaxy model, a hydrostatic model of star-forming gas,
and a dynamical model for a dust grain. We target
high-redshift galaxies which experience the first major star
formation in a gas-rich environment.
%%$z>5$, down to a slightly lower redshift than the highest redshifts of ALMA-detected
%%galaxies ($z\sim 7$) to cover the sufficient redshift range for the possible first
%%dust enrichment.
For simplicity, we neglect the change of grain radius by sputtering.
This is valid for most of the virial mass range of interest, since the
virial temperature is lower than $10^6$ K.\footnote{Thermal sputtering is negligible
at temperature $<10^6$ K \citep{Draine:1979aa}. The virial temperature
is estimated in Appendix \ref{app:age} (equation \ref{eq:Tvir}), and is
lower than $10^6$ K for $M_\mathrm{vir}\lesssim 10^{11}$ M$_{\sun}$.
As shown later, galaxies with $M_\mathrm{vir}\sim 10^{10}$--$10^{11}$ M$_{\sun}$
are important for grain escape. Also, objects with $M_\mathrm{vir}>10^{11}$ M$_{\sun}$
are rare at $z\gtrsim 10$ \citep[e.g.][]{Reed:2003aa}. As we show later,
the grain velocity is less than 100 km s$^{-1}$ in the height ($\zeta$ introduced later)
range of interest
($\lesssim$ several tens of pc); thus, nonthermal sputtering can be neglected as well.}

We treat a dust grain as a test particle, and consider its
motion equation. {In this paper, we consider silicate as
a representative dust species unless otherwise stated, but we also consider graphite to
examine the material dependence of the grain motion.}
The grain is assumed to be spherical with a
uniform material density denoted as $s$. The grain mass $m$ is
estimated as
\begin{align}
m=\frac{4}{3}\upi a^3s.
\end{align}
We give the grain radius as a free parameter.
We adopt $s=3.5$ and 2.24 g cm$^{-3}$ \citep{Weingartner:2001aa}
for silicate and graphite, respectively.
%%We assume silicate for the grain species unless otherwise stated.
We fix the gas structure; that is, we neglect the feedback caused by
dust motion through gas drag.
This is justified since the stellar surface brightnesses
(densities) of galaxies we consider in this
paper are much lower than the value at which the radiation pressure affects the gas structure
\citep{Crocker:2018aa}.

\subsection{Galaxy model}\label{subsec:galaxy}

Here we provide the basic galaxy structures based on which
we estimate relevant forces (gravity, gas drag, and radiation pressure).
The gravitational potential is broadly determined by
the matter density as a result of the cosmological
structure formation. For gas drag, we need to model a
gravitational equilibrium structure of the gas component.
For radiation pressure,
we consider the luminosity of stars formed from the gas
component. We explain the galaxy model in what
follows.

We consider a galaxy with virial mass $M_\mathrm{vir}$ formed
at redshift $z_\mathrm{vir}$ (i.e.\ $M_\mathrm{vir}$ and $z_\mathrm{vir}$
are given parameters). We expect that this object has a
virial radius of $r_\mathrm{vir}$ described by \citep[e.g.][]{Kitayama:1996aa}
\begin{align}
\frac{4}{3}\pi r_\mathrm{vir}^3\Delta_\mathrm{c}\rho_\mathrm{c0}\Omega_\mathrm{M}
(1+z_\mathrm{vir})^3=M_\mathrm{vir},\label{eq:rvir}
\end{align}
where $\Delta_\mathrm{c}$ is the overdensity of a gravitationally bound object
normalized to the mean cosmic matter density, and
$\rho_\mathrm{c0}\equiv 3H_0^2/(8\pi G)$ is the critical
density of the Universe at $z=0$ ($G$ is the gravitational constant).
We adopt $\Delta_\mathrm{c}=200$
\citep[e.g.][]{Peebles:1980aa,Kitayama:1996aa}.
Because of the dissipative nature, 
the cooled portion of the gas in the galaxy halo
is expected to be settled in a disc whose radius
{denoted as $r_\mathrm{disc}$}
is determined by the
initial angular momentum \citep[e.g.][]{Ferrara:2000aa}. In this situation, the
typical surface density of the baryon in the disc ($\Sigma_\mathrm{b}$)
is estimated as
\begin{align}
\Sigma_\mathrm{b}=
\frac{(\Omega_\mathrm{b}/\Omega_\mathrm{M})f_\mathrm{disc}M_\mathrm{vir}}
{\pi q_\mathrm{disc}^2r_\mathrm{vir}^2},\label{eq:Sigma_b}
\end{align}
where $q_\mathrm{disc}\equiv r_\mathrm{disc}/r_\mathrm{vir}$ is the disc radius
normalized to the virial radius,
$f_\mathrm{disc}$ is the fraction of baryon contained in the
disc [fraction ($1-f_\mathrm{disc}$) is contained in the halo;
see equation \ref{eq:rho_halo}] and the baryon fraction in the galaxy is assumed
to be equal to the cosmic
mean ($\Omega_\mathrm{b}/\Omega_\mathrm{M}$).
Considering
the conservation of angular momentum with a typical value of
the spin parameter ($0.04$; \citealt{Barnes:1987aa}),
we obtain
$q_\mathrm{disc}\simeq 0.18$
\citep{Ferrara:2000aa}.
Note that we give $M_\mathrm{vir}$ and $z_\mathrm{vir}$ in our model,
and  eliminate $r_\mathrm{vir}$ in equation (\ref{eq:Sigma_b})
using equation (\ref{eq:rvir}). As a consequence, we
obtain $\Sigma_\mathrm{b}$ for a given set of $M_\mathrm{vir}$ and
$z_\mathrm{vir}$. We fix $q_\mathrm{disc}=0.18$ unless otherwise stated,
but vary $f_\mathrm{disc}$ later.
{For reference, we give the numerical estimates of $r_\mathrm{vir}$ and
$r_\mathrm{disc}$ in equations and (\ref{eq:rvir_num}) and (\ref{eq:rdisc}),
respectively, of Appendix \ref{app:age}.}
%%\begin{align}
%%\Sigma_\mathrm{b}=
%%\frac{4f_\mathrm{disc}r_\mathrm{vir}\Delta_\mathrm{c}\rho_\mathrm{c0}
%%\Omega_\mathrm{b}(1+z_\mathrm{vir})^3}{3q_\mathrm{disc}^2}.\label{eq:Sigmab}
%%\end{align}
The uncertainty caused by the assumed geometry of gas and stars
can be examined practically by changing $q_\mathrm{disc}$. For example, if the gas
and stars are spherically distributed in a compact manner, this case is
approximated by a small $q_\mathrm{disc}$. However, the variation of
$q_\mathrm{disc}$ does not affect our conclusion (Section \ref{subsec:uncertainties}).

We concentrate on the physical
properties perpendicular to the disc, since we are interested
in the dust motion toward the galaxy halo. In this way, the
problem is reduced to one dimension.
%%In reality, various stellar feedback mechanisms will create complicated
%%structures on a short time-scale.
We expect that
our treatment gives a representative estimate even in the presence
of complicated structures as long as we are interested in the
dust motion from the ISM (i.e.\ a concentration of
the baryonic component) to the outer direction.
We leave the complexity arising from multi-dimensional structures
for future work.

We consider stars as radiation sources. We introduce
the star formation efficiency $\epsilon_\star$, which expresses the
fraction of gas converted into stars. With this quantity,
the surface densities of stars ($\Sigma_\star$) and gas ($\Sigma_\mathrm{gas}$)
are written, respectively, as
\begin{align}
\Sigma_\star =\epsilon_\star\Sigma_\mathrm{b},~~~\mbox{and}~~~
\Sigma_\mathrm{gas}=(1-\epsilon_\star )\Sigma_\mathrm{b}.
\label{eq:Sigma_star}
\end{align}
Precisely speaking, a part of $\Sigma_\star$ has been returned into the
gas; thus, $\Sigma_\star$ is not exactly the total mass of the surviving stars.
We should keep in mind that our definition of the stellar mass is the integral of
the past star formation rate (SFR).
In solving radiation transfer and dynamics below, we assume homogeneity
in the directions parallel to the disc plane.
We use coordinate $\zeta$ in this direction with $\zeta =0$ corresponding to the
disc midplane. We also approximate that the
homogeneous disc extends to the infinity. This assumption is equivalent to
$\zeta\ll r_\mathrm{disc}$, which holds for our calculations.
%%(see equation \ref{eq:rdisc} in Appendix \ref{app:age} for the numerical estimate
%%of $r_\mathrm{disc}$).

%%In summary, the parameters characterizing a galaxy are
%%$M_\mathrm{vir}$, $z_\mathrm{vir}$, $f_\mathrm{disc}$, and
%%$\epsilon_\star$ in our simple model. We fix $q_\mathrm{disc}=0.18$ unless
%%otherwise stated.

\subsection{Gravity}\label{subsec:gravity}

For the gravitational field, we need to consider the
contributions from both baryon and dark matter.
Here we formulate the gravitational field effectively including
these two components
based on the formula applied to the Galaxy. Since we
are only interested in the vertical direction of the baryonic
disc, we employ the following functional form for
the gravitational field \citep{Franco:1991aa}:
\begin{align}
g_\zeta=\beta\tanh\left(\frac{\zeta}{H_\mathrm{M}}\right) ,
\end{align}
%%\texttt{Note: I just followed Franco's notation for $g_\zeta$, and it indicates
%%also that the direction of gravity is in the $\zeta$ direction.}
where
$\beta =4\pi GH_\mathrm{M}\rho_\mathrm{tot}(0)$ [$\rho_\mathrm{tot}(0)$ is the total mass
density at the midplane], and $H_\mathrm{M}$ is the mass scale height.
This function is valid for an isothermal self-gravitating disc.
\citet{Franco:1991aa} used this gravity model for the structure of the Galactic disc
\citep[see also][]{Ferrara:1993aa}, and we adopt it for the sake of simplicity.
This expression is still modified
to include appropriate scaling of $H_\mathrm{M}$
and $\beta$ with virial mass and redshift as we explain below.

Since $H_\mathrm{M}\rho_\mathrm{tot}(0)$ roughly gives the matter surface density,
we expect that it has the same scaling as
$M_\mathrm{vir}/r_\mathrm{vir}^2\propto M_\mathrm{vir}^{1/3}(1+z_\mathrm{vir})^{2}$
[note that $r_\mathrm{vir}\propto M_\mathrm{vir}^{1/3}(1+z_\mathrm{vir})^{-1}$ from
equation (\ref{eq:rvir}), and recall that we give constant parameters
$M_\mathrm{vir}$ and $z_\mathrm{vir}$]. Therefore, we assume that
$\beta\propto M_\mathrm{vir}^{1/3}(1+z_\mathrm{vir})^{2}$. This scaling reflects
the tendency of stronger gravitational field in higher-redshift
galaxies as a result of higher matter density.
Assuming $\rho_\mathrm{tot}(0)\propto (1+z_\mathrm{vir})^{3}$, we obtain
$H_\mathrm{M}\propto\beta /\rho_\mathrm{tot}(0)\propto
M_\mathrm{vir}^{1/3}(1+z_\mathrm{vir})^{-1}$.
Based on the above scaling relations, we adopt the following expressions
for $\beta$ and $H_\mathrm{M}$:
\begin{align}
\beta &= \beta_\odot (M_\mathrm{vir}/10^{12}~M_\odot)^{1/3}(1+z_\mathrm{vir})^{2},
\label{eq:beta}\\
H_\mathrm{M} &= H_\mathrm{M\odot}(M_\mathrm{vir}/10^{12}~M_\odot)^{1/3}
(1+z_\mathrm{vir})^{-1},\label{eq:H_M}
\end{align}
where the solar neighbourhood values
($\beta_\odot$ and $H_\mathrm{M\odot}$) are used to give local
calibrations. We adopt $\beta_\odot =3.07\times 10^{-9}$ cm s$^{-2}$
%%betasol=3.074e-9
and $H_\mathrm{M\odot}=250$ pc
\citep{de-Boer:1991aa,Ferrara:1993aa}.
In the normalization, we assume that the Milky Way has established the
current virial mass ($\sim 10^{12}~M_\odot$; \citealt{Trimble:2000aa})
at $z\sim 0$.
In reality, the density structure within the galaxy is
complex because of non-spherical distribution of dark matter
and complicated structure formation of gas and stars.
Although there are uncertainties
in the values of $\beta$ and $H_\mathrm{M}$, the above simple
scaling relations (as a function of $z_\mathrm{vir}$ and $M_\mathrm{vir}$)
are useful to examine the systematic difference
in gravitational potential from the preceding work in the solar
neighbourhood.

\subsection{Hydrostatic equilibrium of the gas}\label{subsec:hydrostatic}

The density stratification of gas is important for the grain motion,
since it affects drag and extinction as a function of $\zeta$. The gas density
in the disc, $\rho_\mathrm{disc}(\zeta )$, is determined by considering the hydrostatic condition:
\begin{align}
\sigma^2\frac{\mathrm{d}\rho_\mathrm{disc}(\zeta )}{\mathrm{d}\zeta} %%-f_\mathrm{g,rad}
+\rho_\mathrm{disc}(\zeta )g_\zeta =0,
\label{eq:hydrostatic}
\end{align}
where $\sigma$ denotes the effective sound speed
including the turbulent velocity.
%%(i.e.\ we denote the equation of state as $P_\mathrm{g}=\rho\sigma^2$,
%%where $P_\mathrm{g}$ is the gas pressure).
For simplicity, we
assume $\sigma$ to be constant. We solve this equation with a
boundary condition of $\rho_\mathrm{disc}(\zeta =0)\equiv\rho_0$ and adjust $\rho_0$
to satisfy the total column density as
\begin{align}
\int_0^\infty\rho_\mathrm{disc}(\zeta )d\zeta =\Sigma_\mathrm{gas}.\label{eq:column}
\end{align}
%%In reality, we solve equations (\ref{eq:hydrostatic}) and (\ref{eq:column})
%%in an iterative way. The obtained solution for $\rho$ is denoted as
%%$\rho_\mathrm{disc}(\zeta )$.

The density declines exponentially at large $\zeta$.
In reality, we expect that the gas density at large $\zeta$ approaches
to a value expected for the gaseous halo. %%Since we are only interested in
%%the interface between the disc and the halo,
We use the following mean gas
density $\rho_\mathrm{g,halo}$ for the halo:
\begin{align}
\rho_\mathrm{g,halo}=\frac{(1-f_\mathrm{disc})M_\mathrm{vir}
\Omega_\mathrm{b}/\Omega_\mathrm{M}}
{\frac{4}{3}\pi r_\mathrm{vir}^3}.\label{eq:rho_halo}
\end{align}
The final gas density profile is determined by
\begin{align}
\rho (\zeta )=\max\left[\rho_\mathrm{disc}(\zeta ),\,\rho_\mathrm{g,halo}
\right] .
\end{align}
%%For presentation, it is sometimes useful to show the number density.
We also use the hydrogen number density, $n_\mathrm{H}$, which
is related to $\rho$ as
\begin{align}
n_\mathrm{H}=\rho /(\mu_\mathrm{H}m_\mathrm{H}),
\end{align}
where $\mu_\mathrm{H}(=1.4)$ is the gas mass per hydrogen, and
$m_\mathrm{H}$ is the mass of hydrogen atom.
The velocity dispersion $\sigma$ is a given parameter.
Unless otherwise stated, we assume that $\sigma =10$ km s$^{-1}$, which is
the typical velocity dispersion of the warm neutral medium in the Galactic disc
\citep[e.g.][]{Spitzer:1978aa} and in nearby galaxies \citep[e.g.][]{Young:1997aa}.

For convenience, we define the typical scale height,
$H_\mathrm{g0}$, as (see also equation \ref{eq:beta})
\begin{align}
H_\mathrm{g0}\equiv\frac{\sigma^2}{\beta}
\simeq 4.0\left(\frac{M_\mathrm{vir}}{10^{10}~\mathrm{M}_{\sun}}\right)^{-1/3}
\left(\frac{1+z_\mathrm{vir}}{11}\right)^{-2}~\mathrm{pc}.\label{eq:scale}
%%4.047
\end{align}
This is much thinner than the discs of the Galaxy and nearby galaxies
\citep[e.g.][]{Yim:2014aa,Nakanishi:2016aa}, which is
a consequence of the redshift dependence of the gravity
(recall that higher densities at higher redshifts lead to stronger
gravity). In reality, the disc structure would be disturbed by H \textsc{ii}
regions and SNe, but the above static dense disc would give
strong gas drag (and a conservative estimate of
grain escape from the disc).
We also define the exponential height as $\zeta$ at which
the density drops $1/\mathrm{e}$ times the central value.
%%This is roughly $\sim 2H_\mathrm{g0}$.
Because the gravity is weak at
$\zeta\ll H_\mathrm{g0}$, the actual scale height (exponential height)
is larger than $H_\mathrm{g0}$ by a factor of $\sim 2$.

\subsection{Radiation pressure}

For the radiation field, we consider the emission from stars.
To calculate the spectral energy distribution (SED) of the
stellar component, we adopt \citet{Bruzual:2003aa}. The SED per stellar
mass ($\ell_\nu$) is calculated with a constant SFR
with a duration (age) of $t_\star$.
To simplify the computation, we fix the stellar SED
(do not vary the SED along time $t$). The calculation is
valid for $t\lesssim t_\star$ (otherwise, the time is contradictory with the
stellar age).
Instead of modeling the complication in the past star formation
history, we vary $t_\star$.
For a test, we also examined the instantaneous burst
with an age of $t_\star$, but we found that radiation pressure is
too weak to push the grains to the halo if $t_\star\gtrsim 30$ Myr.
%%For $t_\star\lesssim 10$ Myr, the instantaneous burst and the continuous
%%star formation produce similar results.
Therefore, we confirmed that an ongoing star formation activity is
essential for radiation pressure to work effectively.
The stellar metallicity is assumed to be 0.004 ($\sim 1/5$~Z$_\odot$;
considering low metallicity at high redshift)
and the Chabrier initial mass function \citep{Chabrier:2003aa}
with a stellar mass range of 0.1--100 M$_{\sun}$ is adopted.
There are some degeneracies among the parameters, but the
age has a largest impact on the SED. Thus, we only
move $t_\star$ for the stellar properties in this paper.

We estimate the stellar surface brightness SED as a function of
frequency $\nu$ (denoted as $\Sigma_{\star\nu}$) using the stellar
surface density in equation (\ref{eq:Sigma_star}):
\begin{align}
\Sigma_{\star\nu}=\Sigma_\star\ell_\nu .
\end{align}
Recall that $\ell_\nu$ is the SED per stellar mass calculated
above. For simplicity, we assume that the stars are
concentrated in the midplane (i.e., at $\zeta =0$).
This assumption does not affect our conclusion since we are
mainly interested in the grain motion at higher $\zeta$ than
the scale height. In this case,
the radiation transfer equation for the intensity
$I_\nu =I_\nu (\zeta ,\,\mu )$ is written using the gas density profile
$\rho (\zeta )$ in Section \ref{subsec:hydrostatic} as
\begin{align}
\mu\frac{dI_\nu}{d\zeta}=-\kappa_\mathrm{g,abs}(\nu)\rho (\zeta )
I_\nu+\frac{1}{4\pi}\Sigma_{\star\nu}\delta (\zeta ),
\end{align}
where $\mu =\cos\theta$ ($\theta$ is the angle from the $\zeta$
direction), $\kappa_\mathrm{g,abs}(\nu )$ is
the mass absorption coefficient of the gas (the opacity is contributed
from the dust) at frequency $\nu$, and $\delta (\zeta )$ is Dirac's
delta function. For simplicity, we neglect scattering.
%%we treat scattering as absorption.
This could cause an underestimate of dust optical depth
by a factor of 2 or less; however, the difference in the grain opacity caused by
various grain size distributions has a comparable uncertainty. Moreover,
we consider a large range for the dust abundance, for which we do not discuss
the precision within factor 3. Therefore, neglecting scattering does not
affect our conclusions in this paper.
By solving this equation, we obtain
\begin{align}
I_\nu (\zeta ,\,\mu )=\frac{\Sigma_{\star\nu}}{4\pi\mu}\exp\left(
-\frac{1}{\mu}\int_0^\zeta\kappa_\mathrm{g,abs}(\nu )\rho (\zeta' )\,\mathrm{d}\zeta'\right) .
\end{align}
%%The radiation force per gas mass, $f_\mathrm{g,rad}(\zeta )$, is estimated as
%%\begin{align}
%%f_\mathrm{g,rad}(\zeta )=\frac{2\pi}{c}\int_0^\infty d\nu\int_0^1d\mu\,
%%\kappa_\mathrm{g,rad}(\nu )I_\nu (\zeta ,\,\mu )\mu^2,
%%\end{align}
%%where $\kappa_\mathrm{g,rad}(\nu )$ is the effective mass absorption
%%coefficient for radiation pressure.

The mass absorption coefficient %% above two coefficients for radiation,
$\kappa_\mathrm{g,abs}(\nu )$ is %%and $\kappa_\mathrm{g,rad}(\nu )$, are
estimated as
\begin{align}
\kappa_\mathrm{g,abs}(\nu )=\mathcal{D}\,
\frac{\int_0^\infty\pi a^2Q_\mathrm{abs}(a,\,\nu )n(a)\, da}
{\int_0^\infty\frac{4}{3}\pi a^3sn(a)\, da},
\end{align}
%%\begin{align}
%%\kappa_\mathrm{g,rad}(\nu )=\mathcal{D}\,
%%\frac{\int_0^\infty\pi a^2Q_\mathrm{rad}(a,\,\nu )n(a)\, da}
%%{\int_0^\infty\frac{4}{3}\pi a^3sn(a)\, da},
%%\end{align}
where $\mathcal{D}$ is the dust-to-gas mass ratio (hereafter simply referred to as the dust-to-gas ratio),
%%$a$ is the grain radius (we assume grains to be spherical throughout
%%this paper),
$Q_\mathrm{abs}(a,\,\nu )$ is %%and $Q_\mathrm{rad}(a,\,\nu )$
the absorption cross-section normalized to the geometrical cross-section.
%%[composed of the absorption and scattering parts
%%as $Q_\mathrm{ext}=Q_\mathrm{abs}+Q_\mathrm{sca}$],
%%$s$ is the grain material density,
and $n(a)$ is the grain size distribution [$n(a)\,\mathrm{d}a$ is proportional to the number
of grains with radii between $a$ and $a+\mathrm{d}a$; the normalization of $n(a)$ cancels out
in the above expressions].
The grain size distribution is assumed to be described by a power-law
suggested by \citet[][hereafter MRN]{Mathis:1977aa}:
\begin{align}
n(a)=\left\{
\begin{array}{ll}
Ca^{-3.5} & \mbox{if $a_\mathrm{min}<a<a_\mathrm{max}$,} \\
0 & \mbox{otherwise},\\
\end{array}
\right.
\end{align}
where $C$ is the normalization constant, which cancels out as mentioned
above. We adopt $a_\mathrm{min}=0.005~\mu$m and
$a_\mathrm{max}=0.25~\mu$m according to MRN.
The absorption cross-section factor $Q_\mathrm{abs}(a,\,\nu )$
%%and $Q_\mathrm{rad}(a,\,\nu )$, are
is calculated using the Mie theory \citep{Bohren:1983aa} with the same optical
constants as adopted in \citet{Weingartner:2001aa}.
%%We adopt $s=3.5$ g cm$^{-3}$ \citep{Weingartner:2001aa} for silicate.
{We adopt silicate, which is consistent with the extinction curve in
the Small Magellanic Cloud \citep{Pei:1992aa}.}
The changes of the grain size distribution and {the grain species} cause variations in
the grain opacity. %%this is degenerate to the variation of the dust-to-gas ratio.
As shown later, the result is sensitive to the dust-to-gas ratio, while we confirmed that
the grain size distribution and {the grain species} cause a subdominant variation to the results.
For example, the change of $a_\mathrm{min}$ to a larger value (such as 0.1 $\micron$)
causes a drop
of the dust opacity within a factor of 2, which is compensated by the increase of
$\mathcal{D}$ by the same factor (at most factor 2).
We concentrate on the variation of $\mathcal{D}$ in this paper.

%%In fact, the detailed assumptions on the above is not essential for this
%%paper. In fact, if $\mathcal{D}\lesssim 10^{-3}$, the opacity of the
%%galactic disc is so low that the absorption of light is not important
%%for the stellar radiation field. Therefore, the detailed choice of the
%%parameters regarding the dust opacity does not affect our results.
%%The disc opacity becomes significant
%%in the dust-rich regime as $\mathcal{D}\sim 10^{-2}$ regardless of
%%the detailed choices of the above parameters; we are not interested
%%in the detailed grain motion in this phase.

Finally, we estimate the radiation force on the dust grain of interest.
The radiation force as a function of $a$ and $\zeta$,
$F_\mathrm{g,rad}(a,\,\zeta )$, is calculated as
\begin{align}
F_\mathrm{d,rad}(a,\, \zeta )=\frac{2\pi}{c}(\pi a^2)\int_0^\infty\mathrm{d}\nu\int_0^1\mathrm{d}\mu\,
Q_\mathrm{rad}(a,\,\nu )I_\nu (\zeta ,\,\mu )\mu^2,
\end{align}
where $Q_\mathrm{rad}(a,\,\nu )$ is the grain cross-section for radiation
pressure normalized to the geometric cross-section. This quantity is evaluated
as
$Q_\mathrm{rad}(a,\,\nu )=Q_\mathrm{abs}(a,\,\nu )+
(1-g_\mathrm{s})Q_\mathrm{sca}(a,\,\nu )$, where
%%$Q_\mathrm{sca}$ is the scattering cross-section normalized to the geometric
%%cross-section, and
$g_\mathrm{s}$ is the scattering asymmetry factor. The relevant quantities
are calculated by the Mie theory.

{\citet{Weingartner:2001ac} studied other forces
due to photoelectron emission and the photodesorption of adatoms,
which could be important in anisotropic radiation fields.
The force due to photoelectron emission is not efficient if the grain
potential is high: since Coulomb drag caused by highly
positive grain charges is the most important factor in determining grain
escape as shown later, photoelectron emission is not relevant in our
context.
Moreover, this force is at most only comparable to radiation pressure.
Thus, taking photoelectron emission into account does not change our
conclusion significantly. The photodesorption of adatoms could have a
large impact on large ($a\gtrsim 0.1~\micron$) grains, and would help
further to push such large grains toward high $\zeta$.
Thus, our estimates could be conservative for grain escape:
as shown later, grains with $a\sim 0.1~\micron$ escape out of the galactic
disc. For this case, the desorption force makes grain escape even easier.
For grains as large as $a\sim 1~\micron$,
whether or not photodesorption helps grains escape from the disc is
worth investigating, but to do this, further modeling of the physical
conditions in the disc (gas temperature, ionization degree, etc.) is required.
We leave this problem for future work.
}

\subsection{Gas drag}

For the grain motion, gas drag is also important.
We refer to \citet{Draine:1979aa} for the estimate of
gas drag. The drag force, $F_\mathrm{drag}$, is expressed as
\citep[see also][]{McKee:1987aa}
\begin{align}
F_\mathrm{drag}=F_\mathrm{drag,direct}+F_\mathrm{drag,plasma},
\end{align}
where the first and second terms on the right-hand side represent
the effects of direct collisions and Coulomb interaction,
respectively. These two terms are evaluated as
\begin{align}
F_\mathrm{drag,direct}\simeq \pi a^2|v_\zeta |\rho\left(v_\zeta^2+\frac{128}{9\pi}
\frac{k_\mathrm{B}T}{\mu_\mathrm{H}m_\mathrm{H}}\right)^{1/2},
\end{align}
\begin{align}
F_\mathrm{drag,plasma}\simeq 4\pi a^2k_\mathrm{B}T\phi^2\ln\Lambda\sum_i
n_iZ_i^2H(s_i),
\end{align}
where $v_\zeta$ is the grain velocity in the $\zeta$ direction,
$k_\mathrm{B}$ is the Boltzmann constant, $T$ is the gas temperature
(fixed at the beginning of Section~\ref{sec:result}),
%%$\mu_\mathrm{H}$ is the gas mass per hydrogen ($\mu_\mathrm{H}=1.4$),
%%$m_\mathrm{H}$ is the hydrogen mass,
$\phi$ is the grain electric potential energy
normalized to the gas thermal energy (see below), $\ln\Lambda$ is
the Coulomb logarithm, $n_i$ is the number density of ion species $i$
(we consider hydrogen and helium with a number ratio of 9 : 1 with
both species in the first ionized state with an ionized fraction of $f_\mathrm{ion}$),
$Z_i$ is
the charge in units of electron charge ($e$), and $H(s_i)$ is a function
of $s_i$ defined below. The normalized grain potential is estimated by
\begin{align}
\phi =\frac{Z_\mathrm{d}e^2}{ak_\mathrm{B}T},
\end{align}
while the Coulomb logarithm and the function $H$ are, respectively,
given by
\begin{align}
\Lambda=\frac{3}{2ae\phi}\left(\frac{k_\mathrm{B}T}{\pi n_e}\right)^{1/2},
\end{align}
where $n_e$ is the electron number density (we assume
$n_e=1.1f_\mathrm{ion}n_\mathrm{H}$ considering the contribution from helium,
where $f_\mathrm{ion}$ is the ionization degree), and
\begin{align}
H(s_i)\simeq s_i\left(\frac{3}{2}\sqrt{\pi}+2s_i^3\right)^{-1}
\end{align}
with
\begin{align}
s_i\equiv\frac{m_iv_\zeta^2}{2k_\mathrm{B}T}
\end{align}
($m_i$ is the mass of ion species $i$),
following the approximated expression derived by \citet{Draine:1979aa}.
Note that in the above definition, $F_\mathrm{drag}$ is always positive.
Thus, considering the direction,
the drag force is expressed as $-F_\mathrm{drag}v_\zeta /|v_\zeta |$.
We discuss and fix $f_\mathrm{ion}$
at the beginning of Section \ref{sec:result}.

For the grain charge, we consider
the balance between photoelectric charging and collisional charging
following \citet{Inoue:2003ab}.
The collisional charging rate by particle species $i$
(H, He, or e), $R_i$, is
estimated for the Maxwellian velocity distribution
as \citep[e.g.][]{Draine:1987aa}
\begin{align}
R_i=\pi a^2Z_iS_in_i
%%\int_0^\infty\sigma_i(a,\, Z_\mathrm{d},\, Z_i,\, v_i)v_if(v_i)dv_i,
\left(\frac{8k_\mathrm{B}T}{\pi m_i}\right)^{1/2}g(x),
\end{align}
where $S_i$ is the sticking coefficient (assumed to be unity),
%%$m_i$ is the mass of species $i$ (H, He, or e),
$x\equiv Z_i\phi$, and
\begin{align}
g(x)=\left\{
\begin{array}{ll}
1-x & \mbox{for $Z_\mathrm{d}Z_i\leq 0$},\\
\exp (-x) & \mbox{for $Z_\mathrm{d}Z_i>0$}.
\end{array}
\right.
\end{align}
We neglect the image potential, %%keeping in mind that this underestimates
%%the gas drag around zero grain charge \citep{Draine:1987aa}.
since the grains are highly positively charged in the region where the
Coulomb drag force is important.
The photoelectric charging rate, on the other hand, is
given by \citep[e.g.][]{Draine:1978aa}
\begin{align}
R_\mathrm{pe}=\pi a^2\int_0^\infty Q_\mathrm{abs}(a,\,\nu )Y_\nu (a,\, Z_\mathrm{d})
\frac{4\pi J_\nu}{h\nu}d\nu ,
\end{align}
where $Y_\nu$ is the photoelectric yield, $J_\nu$ is the mean
intensity averaged over the solid angle (estimated below),
and $h$ is the Planck constant. We estimate $Y_\nu$ based on
\citet{Weingartner:2001ab}. The mean intensity is calculated as
\begin{align}
J_\nu =\frac{1}{2}\int_0^1I_\nu (\mu )d\mu .
\end{align}
Since the radiation only comes from the lower hemisphere,
the integration range is from 0 to 1 (i.e.\ $I_\nu =0$ for
$-1\leq\mu <0$).

\subsection{Motion equation of a dust grain}

The motion equation of a dust grain is written as
\begin{align}
m\frac{\mathrm{d}v_\zeta}{\mathrm{d}t}=F_\mathrm{d,rad}-mg_\zeta-
F_\mathrm{drag}\frac{v_\zeta}{|v_\zeta |}.\label{eq:motion}
\end{align}
Each term on the right-hand side is evaluated in one of
the previous subsections. We adopt the boundary condition
$v_\zeta =0$ at $\zeta =0$, but this is not important because the
grain velocity at low $\zeta$ is determined by the equilibrium value
($v_\zeta$ such that $\mathrm{d}v_\zeta /\mathrm{d}t=0$).
We comment on the effect of varied initial velocities
in Section \ref{subsec:uncertainties}.

{Since the drag time-scale could be extremely short, we do not
adopt an explicit discretization method for the time.
We discretize the coordinate $\zeta$ to solve equation (\ref{eq:motion})
and determine the grain motion iteratively between the two grid points
(the discrete $i$th grid point is denoted as $\zeta_i$).
The width between the grid points
are set as a few per cent of the typical scale height estimated in
equation (\ref{eq:scale}). From the velocity at $\zeta_i$ to
$\zeta_{i+1}$, we apply an analytic solution based on the solution at
$\zeta_i$ and update the grain velocity at $\zeta_{i+1}$. This
velocity is used for the next iteration. We repeat this until the velocity
at $\zeta_{i+1}$ converges.
This iterative procedure stabilizes the obtained
solution under a fixed spatial grid.
}

\subsection{Varied parameters}\label{subsec:list_param}

There are some free parameters whose values should be specified.
The parameters are categorized into two types: one is
related to galaxies, and the other to dust. In what follows,
we explain the range of the parameter values we adopt in this
paper. The varied parameters are summarized in Table \ref{tab:param}.

\begin{table}
\caption{Parameters.}
\begin{center}
\begin{tabular}{lcccc}
\hline
Parameter & units & fiducial value & minimum & maximum \\
\hline
$M_\mathrm{vir}$ & $M_{\sun}$ & $10^{10}$ & $10^8$ & $10^{12}$ \\
$z_\mathrm{vir}$ & & 10 & 6 & 14\\
$\epsilon_\star$ & & 0.3 & 0.1 & 0.5\\
$t_\star$ & Myr & 30 & 10 & 100 \\
$a$ & $\micron$ & 0.1 & 0.01 & 1\\
$\mathcal{D}$ & & $10^{-3}$ & $10^{-4}$ & $10^{-2}$\\
\hline
\end{tabular}
\label{tab:param}
\end{center}
\end{table}

For the galaxy model, we move $M_\mathrm{vir}$, $z_\mathrm{vir}$,
$\epsilon_\star$, and $t_\star$. We are interested in the dust enrichment at high redshift
$z\gtrsim 7$, where we already know that dusty galaxies existed as mentioned
in the Introduction; thus, we examine $z_\mathrm{vir}\sim 10$
($z_\mathrm{vir}=6$--14). We consider $M_\mathrm{vir}=10^8$--$10^{12}$ M$_{\sun}$,
which covers the virial masses of the objects contributing to
high-redshift star formation activities \citep[e.g.][]{Ciardi:2000aa}.
For the star formation efficiency, we survey a range of
$\epsilon_\star=0.1$--0.5. As we find below, if
$\epsilon_\star \leq 0.1$, dust is not pushed efficiently by radiation pressure.
For $t_\star$, we examine 10, 30, and 100 Myr.
We fix $f_\mathrm{disc}=0.5$ because of the degeneracy with
$\epsilon_\star$. Note that $f_\mathrm{disc}\epsilon_\star$ is the
fraction of the total baryonic mass converted into stars.

For the grain parameters, we vary $a$ and $\mathcal{D}$.
We consider a range of $a=0.01$--1 $\micron$, which roughly covers the grain sizes
in the ISM (e.g.\ MRN). For $a\lesssim 0.01~\micron$, as shown
later, the grains
are effectively trapped in the galactic disc because of inefficient absorption and
scattering of ultraviolet (UV) radiation (i.e.\ weak radiation pressure).
We examine
$\mathcal{D}=10^{-4}$--$10^{-2}$ (corresponding to
$\sim$0.01--1 times the Galactic dust-to-gas ratio).

\section{Results}\label{sec:result}

%%\subsection{Hydrostatic gas profile}\label{subsec:profile}

%%Before we examine the dust motion, we show the hydrostatic structure
%%of the gas to see the effect of radiation pressure on the
%%gas profile. Overall, we find that radiation pressure has little
%%influence on the gas structure. The effect of radiation pressure is
%%larger for less gas surface density, higher star formation efficiency,
%%younger stellar age, and higher dust-to-gas ratio. Therefore,
%%we choose $f_\mathrm{disc}=1$ and $\epsilon_\star =0.1$ (i.e., the
%%resulting stellar-to-gas mass ratio is 9:1). For comparison,
%%we also examine a case without stellar radiation and with the same
%%gas mass ($f_\mathrm{disc}=0.1$ and $\epsilon_\star =0$).
%%We consider a galaxy with $z_\mathrm{vir}=7$ and $M_\mathrm{vir}=10^{10}~M_\odot$.
%For the other parameters, we use the fiducial values.

%%In Fig.\ \ref{fig:hydrostatic}, we show the density profile.
%%Even in the above extreme two cases, the drop of the central
%%density by the radiation pressure is only a factor of 1.5.
%%The radiation pressure does not change the height above which
%%the density drops exponentially. Therefore, we conclude that
%%radiation pressure does not affect the gas profile.

In this section, we examine the grain motion,
focusing on the effects of the parameters in Section \ref{subsec:list_param}.
In the following subsections, we move only one parameter with
the others fixed to the fiducial values shown in Table \ref{tab:param}.
We neglect the time difference between the halo virialization and
the ongoing star formation activity; therefore, we regard the formation
redshift as the same as the observed redshift (i.e.\ $z\simeq z_\mathrm{vir}$).
This could be justified because the duration of star formation
($t_\star\leq 10^8$ yr) is
significantly shorter than the cosmic age (equation \ref{eq:cosmic_age}
in Appendix \ref{app:age}).

Before we start moving each parameter, we set the ionization degree
$f_\mathrm{ion}$ and the gas temperature $T$, which are
important for grain charging and gas drag.
Determining $f_\mathrm{ion}$ {and $T$} requires
a treatment of internal/external ionizing radiation, shock ionization,
etc., and is beyond the scope of this paper.
{To keep our formulation simple (and to avoid additional complexity
arising from the detailed physical states of the gas), we basically fix
$f_\mathrm{ion}$ and $T$ by introducing some simple assumptions
given below.}
%%Our calculations are based on simple assumptions on $f_\mathrm{ion}$.
We expect that $f_\mathrm{ion}\simeq 1$ at high $\zeta$
because of collisional ionization by the halo gas and/or ionizing radiation from
nearby galaxies. In contrast, $f_\mathrm{ion}\sim 0$ at low
$\zeta$ in the star-forming disc (assuming that H \textsc{ii} regions is
confined in small regions). Therefore, we expect that $f_\mathrm{ion}$
varies from zero in the disc, an intermediate value around the disc--halo interface,
and 1 in the halo.

\begin{figure}
\includegraphics[width=0.45\textwidth]{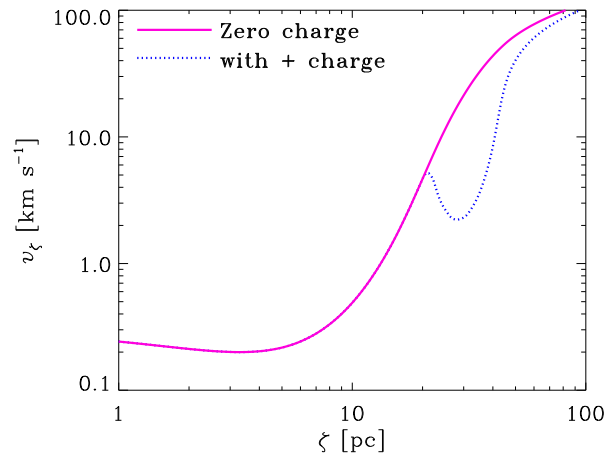}
\caption{Grain velocity $v_\zeta$ as a function of $\zeta$ for the two
test cases regarding the grain charge.
The solid line shows the case 
where $f_\mathrm{ion}=0$.
The dotted line presents the case with $f_\mathrm{ion}=1$. In
this case,
if the resulting grain charge is positive ($Z_\mathrm{d}>0$),
we use this grain charge; otherwise, we assume that
$Z_\mathrm{d}=0$ (this prescription is correct as long as the
gas in the dense part of the disc is neutral).
%%The dashed line shows the case where the grain potential is
%%always $-3$ V. See the text for the explanation for each case.
%%The dot--dashed line displays the same case as the solid line but the
%%gas temperature is changed to $10^5$ K.
\label{fig:vel_charge}}
\end{figure}

In Fig.\ \ref{fig:vel_charge}, we show the grain velocity $v_\zeta$ as a function of $\zeta$
for two different settings regarding $f_\mathrm{ion}$. We adopt the fiducial values for
the parameters (Table \ref{tab:param}).
First, we examine the case where
$f_\mathrm{ion}=0$ for all $\zeta$. In this case, the drag force is only contributed
from direct collisions ($F_\mathrm{drag}=F_\mathrm{drag,direct}$)
regardless of the grain charge.
Second, we examine the case where $f_\mathrm{ion}=1$ for all $\zeta$.
{As mentioned above, this is justified at large $\zeta$.
In the upper part ($\zeta\gg H_\mathrm{g0}$), grains are charged positively.
%%As shown later,  this positive charging is important in strengthening gas drag.
On the other hand, we obtain $Z_\mathrm{d}<0$ for $f_\mathrm{ion}=1$
in the dense part of the disc. However, it is unlikely that the gas in the disc
is all ionized because of its high density ($n_\mathrm{H}\gtrsim 100$ cm$^{-3}$);
rather, the major part of the disc should be maintained neutral for star formation to continue.
Since we do not treat
the detailed physical conditions in the disc, we simply
neglect Coulomb drag if we obtain $Z_\mathrm{d}<0$, which is equivalent to the assumption that the
disc is filled with neutral gas.
%%In any case, this
%%assumption does not affect our results as long as the disc gas is neutral (i.e.\ the Coulomb drag
%%is negligible in neutral gas).
}
For the temperature, because of the efficient cooling in the disc, it is expected that
$T$ is lower than $10^4$ K. 
We assume $T=10^4$ K {in all regions} and discuss the case
with a lower temperature later.
%%Third, we examine the case where $f_\mathrm{ion}=1$
%%and fix the grain potential to $-3$ V, expected for $T_\mathrm{gas}=10^4$ K
%%and no radiation field. This gives an estimate for the maximum drag
%%in the galactic disc (although the assumption is inconsistent with the
%%presence of radiation pressure). Fourth, we present the case with the same setup
%%as the first case but with $T_\mathrm{gas}=10^5$ K. This gives a higher efficiency of
%%gas drag by direct collisions, and more efficient electron attaching (less positive
%%charge).

%%\begin{figure}
%%\includegraphics[width=0.45\textwidth]{hydrostatic.eps}
%%\caption{Hydrostatic structures (hydrogen number density as a
%%function of $\zeta$) for two ``extreme'' models
%%with $(f_\mathrm{disc},\,\epsilon_\star )=(1,\, 0.9)$
%%and (0.1,\,0) with $\mathcal{D}=0.01$ for a galaxy
%%with $z_\mathrm{vir}=7$ and $M_\mathrm{vir}=10^{10}~M_\odot$.
%%We choose the fiducial values for the other parameters.
%%\label{fig:hydrostatic}}
%%\end{figure}

The two cases {discussed in the above paragraph for
$f_\mathrm{ion}=0$ and 1
(and presented in Fig.\ \ref{fig:vel_charge})}
show the same results at $\zeta\lesssim 20$ pc,
since $F_\mathrm{drag}=F_\mathrm{drag,direct}$ in both cases.
%%The third case shows
%%extremely low grain velocities in that $\zeta$ range because
%%gas drag is enhanced by the grain charge (although the grain charge
%%value assumed is extreme). We also observe that at $\zeta\lesssim 20$ pc,
%%if we compare the first two cases, the effect of grain charge is large.
In the second case with $f_\mathrm{ion}=1$, since the electron density drops, the grain is
positively charged by the photoelectric effect at $\zeta\gtrsim 20$ pc.
The positive charge raises gas drag and suppresses the grain
velocity at $\zeta\sim 20$--40 pc. As a consequence, the grain velocity
is much lower in the positively charged case than in the zero charge case.
At $\zeta\gtrsim 40$ pc, because the ion density drops further, the grain
obtains a large velocity.
%%The fourth case provides stronger drag at low $\zeta$ because of higher
%%sound speed, while it gives a weaker drag at high $\zeta$ since more
%%efficient electron attachment makes grains
%%less positively charged.

%%Therefore, we can robustly conclude that the grain obtains a large
%%velocity at large $\zeta$.
%%In the disc, we are interested in a continuously star-forming disc; thus, we
%%expect that the disc is maintained neutral (or the ionization feedback
%%has had a minimum influence of the disc gas). Thus, we assume that the
%%disc is neutral with a temperature of $T_\mathrm{gas}=10^4$ K
%%(with higher temperatures, the disc gas is ionized).
In the above, we assumed that $T=10^4$ K, but the temperature may be
lower at low $\zeta$.
If the temperature is lower, gas drag is weaker. In other words, the
case with $T_\mathrm{gas}=10^4$ K gives the strongest drag
(the most conservative estimate for the grain escape from the disc)
as long
as we assume the galactic disc to be composed of neutral gas.
In contrast, the upper part of the disc
could be ionized (like our Galaxy halo).
Since Coulomb drag dominates the
total drag force, the gas temperature is not important in the
ionized case.

In summary, {the second case with $T=10^4$ K
(the dotted line in Fig.\ \ref{fig:vel_charge})
gives a conservative estimate of the grain velocity} at both low and high $\zeta$.
Therefore, we adopt the second case; that is, we assume that
$f_\mathrm{ion}=1$ but we neglect negative charging, which could occur
at low $\zeta$
(or assume that the disc is neutral in such a region).

\subsection{Effect of grain properties}\label{subsec:grain_param}

\begin{figure*}
\includegraphics[width=0.45\textwidth]{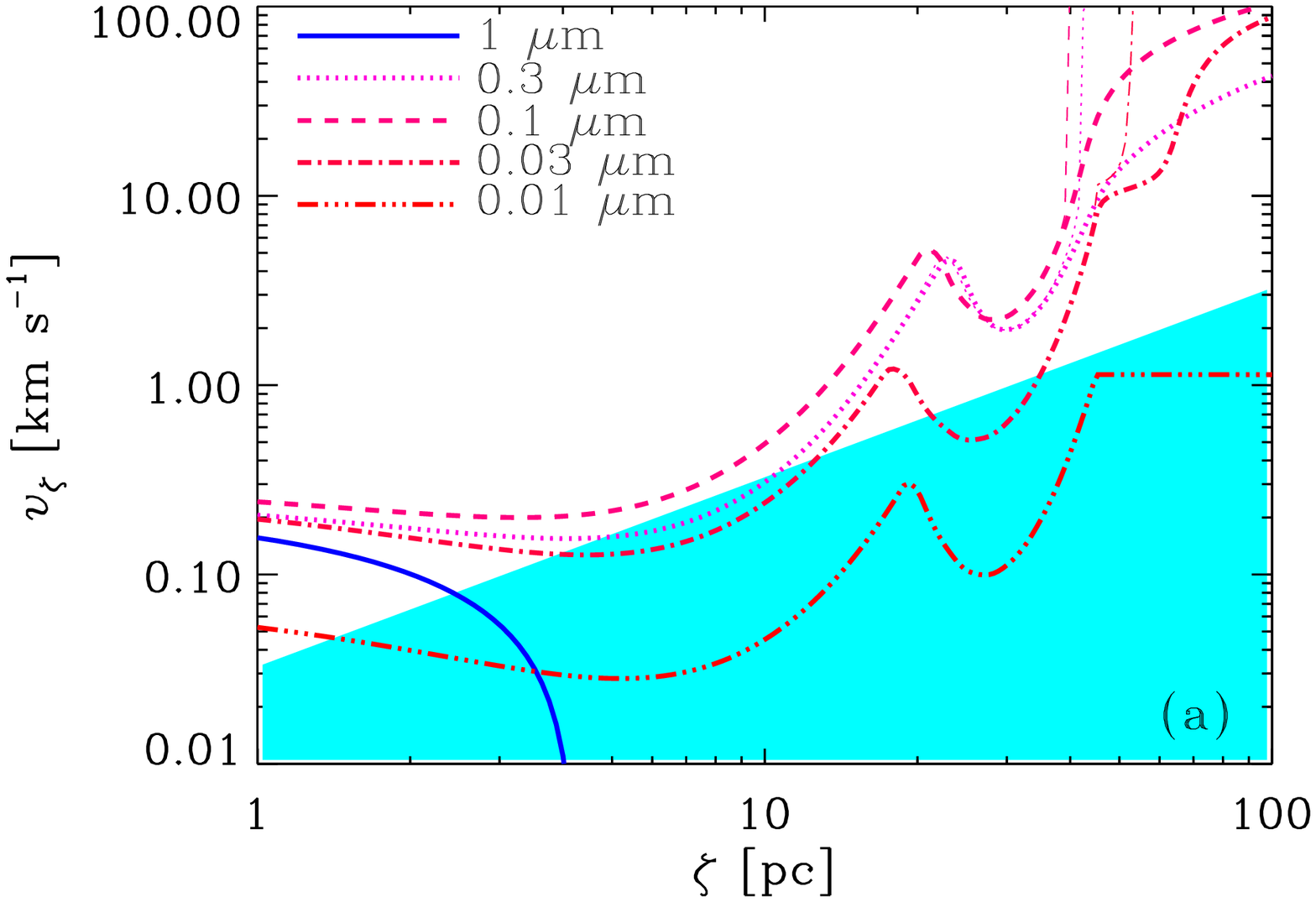}
\includegraphics[width=0.45\textwidth]{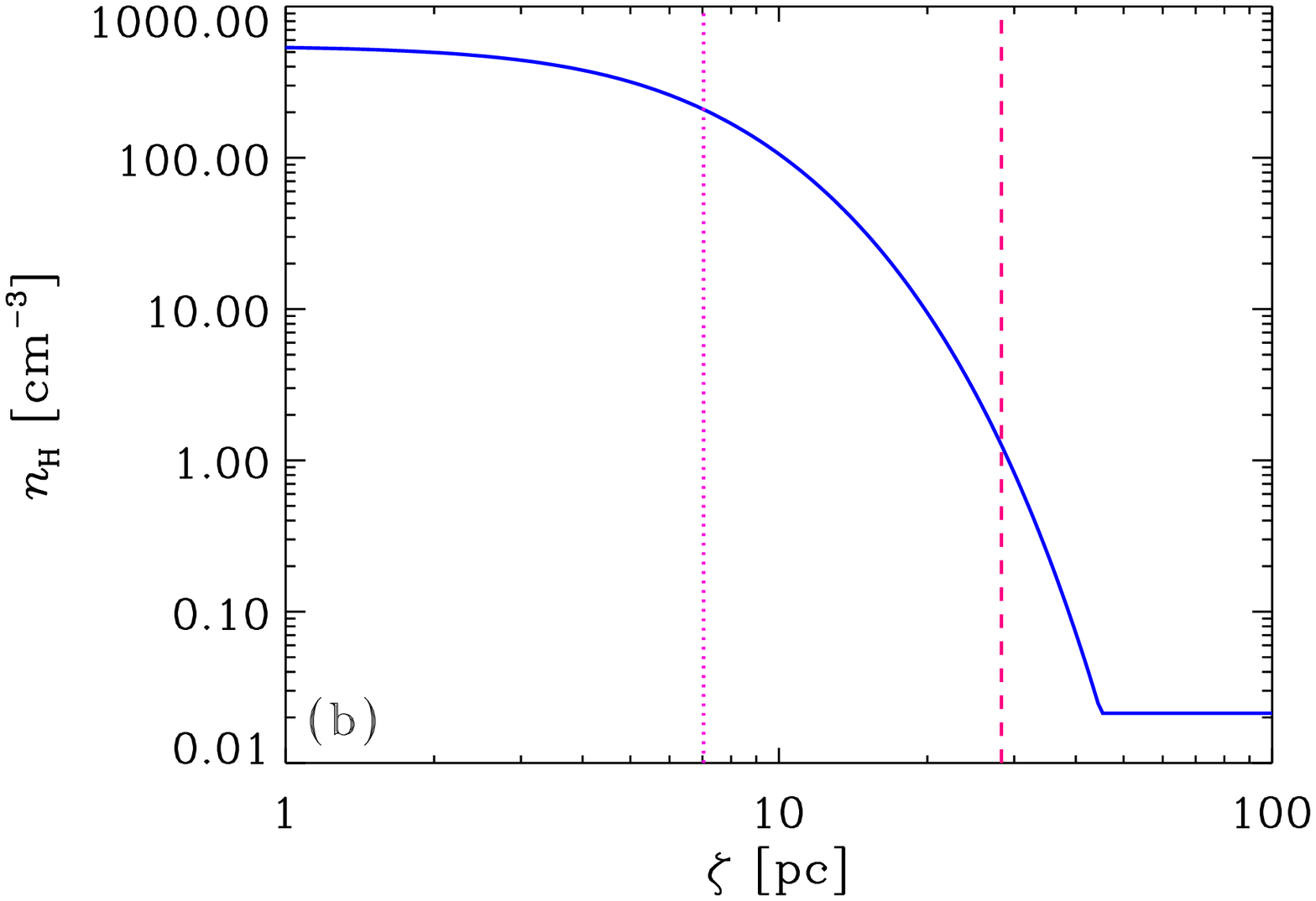}
\includegraphics[width=0.45\textwidth]{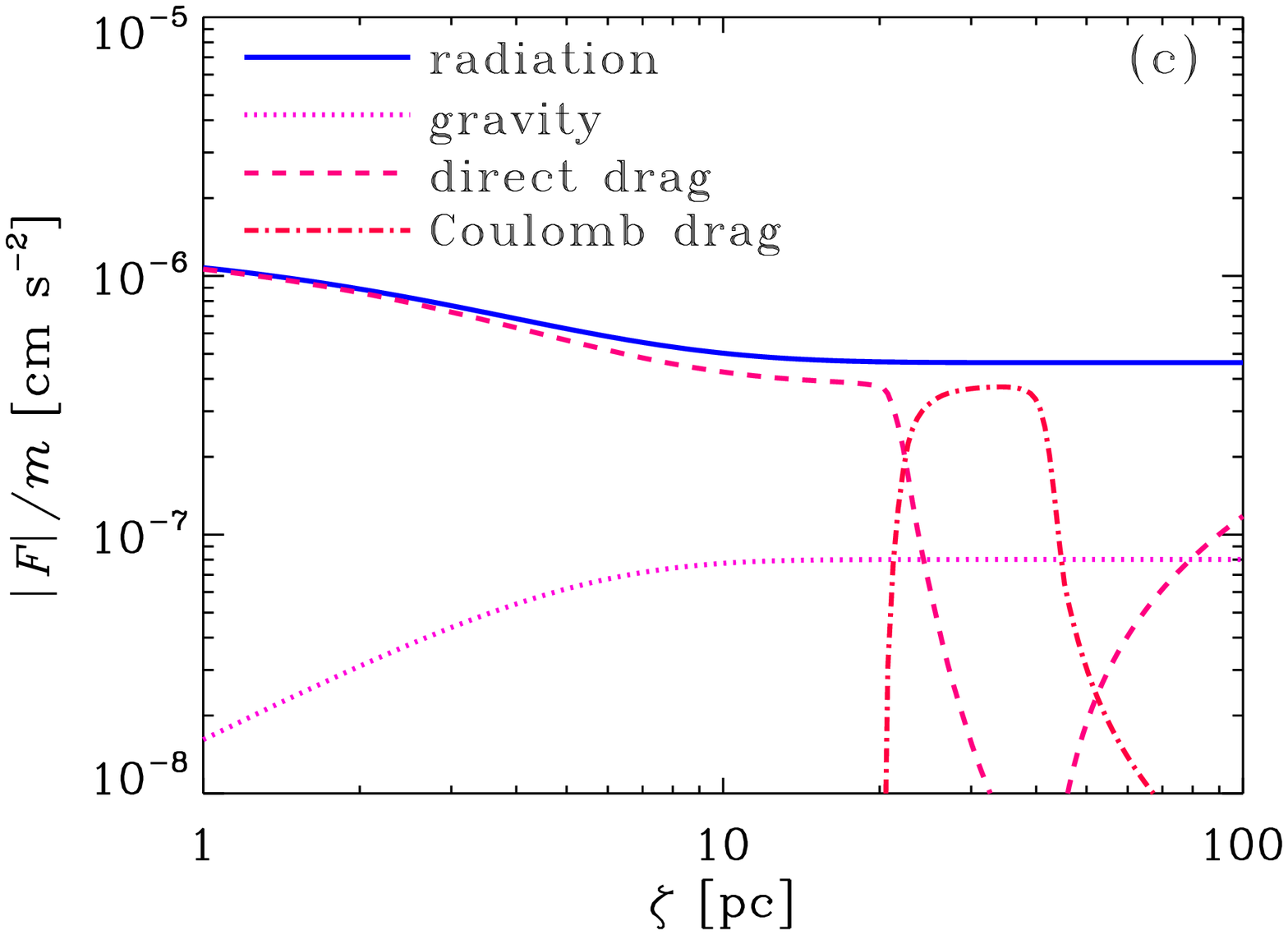}
\includegraphics[width=0.45\textwidth]{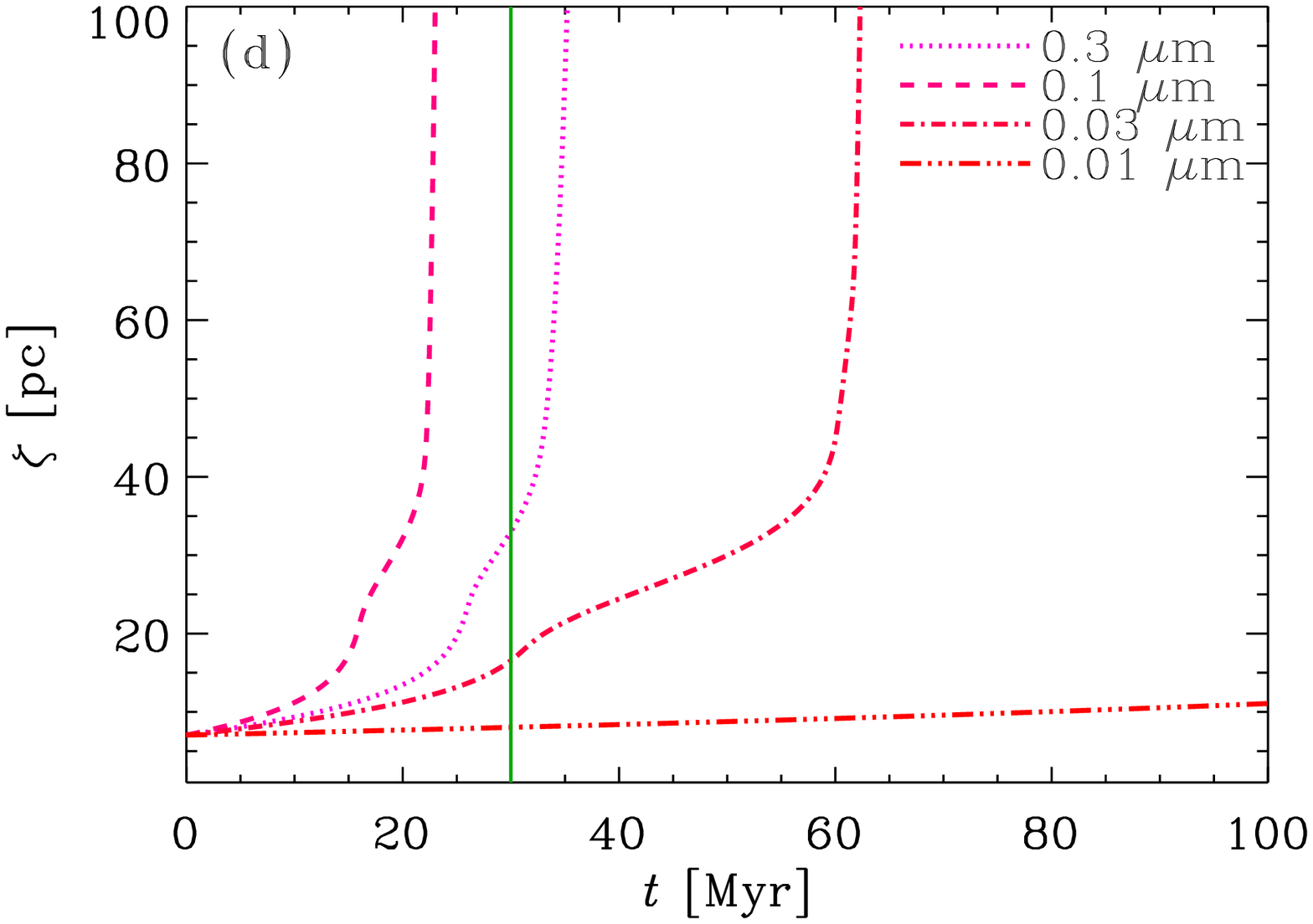}
\caption{
(a) Grain velocity $v_\zeta$ as a function of $\zeta$ for various
grain radii.
The thick solid, dotted, dashed, dot--dashed, and triple-dot--dashed lines
show the results for $a=1$, 0.3, 0.1, 0.03, and 0.01 $\micron$,
respectively. The thin lines, which are almost overlapped with
the thick lines, show the equilibrium velocities (the difference between
the thick and thin lines is seen only at $\zeta\gtrsim 40$ pc for $a=0.03$--0.3 $\micron$).
The shaded region shows $v_\zeta <\zeta/t_\star$, which means
that the grains are effectively trapped in $t_\star$ (duration of the star formation
activity).
(b) Gas profile (hydrogen number density as a function of $\zeta$).
The vertical dotted line shows the exponential height, where the density drops
to the $1/\mathrm{e}$ of the central value ($\zeta =7.0$ pc in this case).
The dashed vertical line presents the height at the local velocity minimum of the $a=0.1~\micron$
grain (due to the trapping by the Coulomb drag; $\zeta =28$ pc in this case; see Panel a).
(c) Contribution from the various forces to the acceleration for $a=0.1~\micron$. The solid,
dotted, dashed, and dot--dashed lines represent the contribution from
radiation force, gravity, drag from direct collisions, and Coulomb drag, respectively.
The absolute values are shown.
Note that only radiation force is in the positive direction while the other
forces are in the negative direction.
(d) Grain position $\zeta$ as a function of time $t$ for various grain radii with the
same line species as in Panel (a).
The position of the grain at $t=0$ is at the exponential height (7.0 pc in this case).
The case of $a=1~\micron$
is not shown here. The vertical line shows $t=t_\star$.
\label{fig:vel_a}}
\end{figure*}

First, we show the dependence on the grain radius.
In Fig.\ \ref{fig:vel_a}a, we present the grain velocity $v_\zeta$ as
a function of $\zeta$ for $a=1$, 0.3, 0.1, 0.03, and 0.01 $\mu$m.
We observe that the velocity is suppressed
to a low level as $v_\zeta\sim 0.1$ km s$^{-1}$ at $\zeta\lesssim 10$ pc.
The largest dust grain
with $a=1~\mu$m once launched is decelerated since gravity becomes dominant over
radiation pressure (recall that gravity scales as $\propto a^3$ while
the radiation pressure increases roughly as $\propto a^2$ for $a\gtrsim 0.1~\micron$).
The grain velocity for $a=0.01~\micron$ is low at low $\zeta$ because it does not
absorb or scatter the stellar light efficiently
{(i.e.\ the efficiency of receiving radiation pressure is low)}.
As mentioned above, because of the grain charge, the grain velocity is
suppressed at $\zeta\sim 30$~pc, although the gas density drops
(see Fig.~\ref{fig:vel_a}b for the gas profile).
The kink structure for $a=0.03$ and 0.01 $\micron$ at $\zeta\sim 40$ pc
corresponds to the sudden transition from the disc density profile to the
halo density. In this sense, we could regard $\zeta\sim 40$ pc as the
disc--halo interface.
At $\zeta\gtrsim 30$ pc,
the grain is accelerated by radiation pressure especially in the cases of
$a=0.1$ and 0.03 $\micron$. In the case of $a=0.01~\micron$, the velocity
approaches the terminal velocity determined by the balance between
gas drag and radiation pressure.
%%(both are constant at large $\zeta$ if the velocity is constant).

We also find that the grain velocity is broadly described by the equilibrium value
at each position. The equilibrium velocity is determined by the solution of
$\mathrm{d}v_\zeta /\mathrm{d}t=0$ in equation (\ref{eq:motion}). In Fig.~\ref{fig:vel_a}a,
we observe that the equilibrium velocity is indistinguishable from the
actual velocity except for the cases of $a=0.03$--0.3 $\micron$ at $\zeta\gtrsim 40$ pc.
Therefore, the grain
velocity is determined by the local condition
for most of the regions
where $v_\zeta\lesssim 10$ km s$^{-1}$.
This is more clearly shown by the individual force components in
Fig.~\ref{fig:vel_a}c (for $a=0.1~\micron$). At $\zeta\lesssim 20$ pc, radiation force is almost
equal to drag from direct collisions, while at $\zeta\sim 20$--40 pc,
Coulomb drag is balanced with radiation force. Gravity is always subdominant for
the escaping grain. Therefore, the grain motion at $\zeta\lesssim 40$ pc is
determined by the balance between radiation force and drag (or in other words, the
velocity is adjusted to achieve this balance). At $\zeta\gtrsim 40$ pc,
the grain is predominantly accelerated by the radiation force. The drag force also increases
at $\zeta\gtrsim 40$ pc because of the increase in grain velocity (note that
the density is constant at such a high $\zeta$; Fig.\ \ref{fig:vel_a}b).

To clarify if the grains are successfully transported outwards within a reasonable
time, we show the `trapped' region, which is defined as
$\zeta /v_\zeta >t_\star$. The trapped region indicates that the crossing time-scale is
longer than the stellar age {after which the radiation pressure stops
acting on the dust grain}.
If the major part of the trajectory of a grain in the $v_\zeta$--$\zeta$
space passes through the trapped region, the grain is effectively trapped by
drag force (or the grain cannot escape within a reasonable time).

We directly show the grain trajectory in the $\zeta$--$t$ space
in Fig.~\ref{fig:vel_a}{d}. We focus on a grain whose initial position is at
the exponential height ($\zeta =7.0$ pc in this case).
If we trace a dust grain at the disc plane, it spends most of its time in
the disc; thus,
we expect that most grains escaping to the halo are located at the upper
($\zeta\sim H_\mathrm{g0}$) area of the disc.
{Indeed, if we start from $\zeta =0$, the grain escape takes
$\sim 50$ Myr for the fiducial parameters. Although the time is still comparable
to $t_\star$, this indicates that grain escape from the mid-plane of the
disc may be difficult. Thus, we `relax' the condition and start from the exponential
height to concentrate on the
grain escape from the upper disc (but still in the dense part).}
Here, the initial velocity is not important since, as shown above, the
grain velocity is determined by the local condition {in the disc}.
We observe in Fig.\ \ref{fig:vel_a}{d} that grains
with $a\sim 0.1~\mu$m are
quickly transported far above the galactic disc, while the `trapped' grains
above (grains larger than 0.3 $\mu$m or smaller than 0.03 $\mu$m)
are not efficiently pushed toward high $\zeta$ within $t_\star$. Only in the case of
$a=0.1~\micron$ is the grain transported to the halo in $t_\star =30$ Myr.
Other grains spend most of the time around $\zeta\sim 20$ pc where the grains
are affected by the strong Coulomb drag.

\begin{figure}
\includegraphics[width=0.45\textwidth]{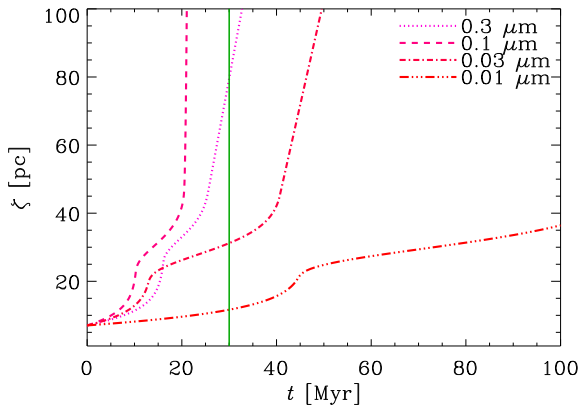}
\caption{Same as Fig.\ \ref{fig:vel_a}{d} but for graphite.
\label{fig:vel_gra_a}}
\end{figure}

The radiation force may depend on the grain material. As pointed out
by \citet{Barsella:1989aa}, graphite grains receive more radiation pressure
than silicates; thus,
graphite may escape from the disc more easily than silicate. On the other hand,
as argued by \citet{Bianchi:2005aa}, graphite grains are charged more easily,
so that they are more trapped by gas (Coulomb) drag.
To investigate the grain material dependence, we show the grain trajectories for
graphite in Fig.\ \ref{fig:vel_gra_a}, which is to be compared with
Fig.\ \ref{fig:vel_a}{d}.
Because graphite is more efficiently pushed by stellar radiation
than silicate, it is more quickly accelerated at low $\zeta$. As
a consequence, the grain with $a=0.3~\micron$
as well as that with $a=0.1~\micron$ is transported to
a high latitude for graphite. At the same time, at $\zeta\gtrsim 40$ pc,
the inclination of the 0.3 $\micron$ grain on the $\zeta$--$t$ diagram
(i.e.\ the grain velocity) is lower for graphite than for silicate. This is
because of stronger Coulomb drag. This result supports both of the above
papers: graphite is easily pushed by radiation pressure in the disc
because of high radiation pressure, while its velocity could be suppressed
around the disc--halo interface by Coulomb drag.
%%Therefore, graphite grains tend to
%%be trapped at the lower layer of the galaxy halo (or in the upper area of
%%the disc) by Coulomb drag. As a result, even in the case of
%%$a=0.1~\micron$, graphite only reaches $\zeta\sim 30$ pc in $t_\star =30$ Myr.
Overall, however, the grain trajectory is not drastically different between
silicate and graphite, so that we argue that the condition of grain escape
is not sensitive to the grain material.

To summarize the above results, grains with $a\sim 0.1~\micron$
successfully escape out of the disc even if we consider strong Coulomb drag
in the upper disc.
Large ($a\gtrsim 1~\micron$) grains are captured by gravity,
while small ($a\lesssim 0.03~\micron$) grains are trapped by gas drag.
Comparing silicate and graphite, there is a slight tendency that
graphite is less trapped by
gas drag in the disc while its velocity is suppressed by Coulomb drag
at the disc--halo interface. Overall, the condition for grain escape is
not sensitive to the grain species. Below, we concentrate on a silicate grain
with $a=0.1~\micron$ unless otherwise stated.

\subsection{Effect of dust-to-gas ratio}\label{subsec:dg}

\begin{figure}
\includegraphics[width=0.45\textwidth]{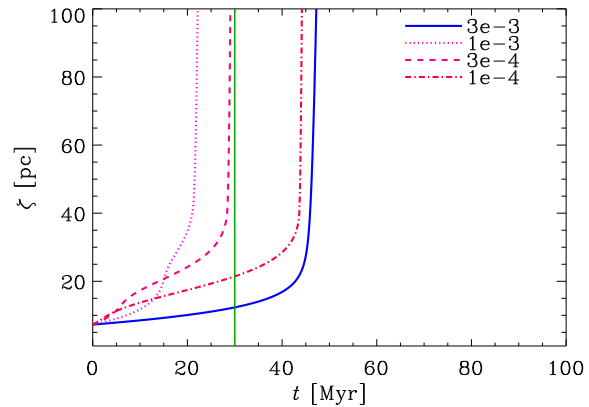}
\caption{Same as Fig.\ \ref{fig:vel_a}{d}, but for various
dust-to-gas ratios {with $a=0.1~\micron$}.
The solid, dotted, dashed, and dot--dashed lines
show the results for $\mathcal{D}=3\times 10^{-3}$, $10^{-3}$, $3\times 10^{-4}$, and
$10^{-4}$, respectively.
\label{fig:vel_dg}}
\end{figure}

The dust abundance (dust-to-gas ratio, $\mathcal{D}$)
affects the radiation pressure through the extinction.
In Fig.\ \ref{fig:vel_dg}, we show the time evolution of the grain position.
%%We observe that, if the dust-to-gas ratio is
%%0.01 (comparable to the Milky Way value), the dust is not pushed
%%effectively by the radiation pressure because of high extinction.
We do not show the case of $\mathcal{D}=0.01$, where the dust velocity is
negative at $\zeta >7.3$ pc because of too strong extinction (too weak radiation
pressure).
On the other hand, if the dust-to-gas ratio is $\lesssim 3\times 10^{-4}$,
it takes more time for the dust grains to reach a high altitude.
This is due to the increased grain charge in the low-extinction cases
(i.e.\ grains are more trapped at $\zeta\sim 20$ pc).
Therefore, there is an `optimum' dust-to-gas ratio $\mathcal{D}\sim 10^{-3}$,
where the stellar light is moderately extinguished to suppress the enhancement
of gas drag by grain charging but still radiation pressure is strong enough.

The optimum dust-to-gas ratio is characterized by the dust extinction
optical depth. Since we are interested in young stellar population,
the major stellar radiation energy absorbed by dust is at UV wavelengths,
typically $\sim 0.2~\micron$ \citep{Buat:1996aa}.
The dust opacity at
0.2 $\micron$ in our model is $\kappa_\mathrm{g,abs}\sim 1.1\times 10^4\mathcal{D}$
cm$^2$ g$^{-1}$. The optical depth at $0.2~\micron$ is estimated
as
$\kappa_\mathrm{g,abs}\Sigma_\mathrm{gas}\simeq
8.3\times 10^{2}\mathcal{D}(1-\epsilon_\star )f_\mathrm{disc}
(M_\mathrm{vir}/10^{10}~\mathrm{M}_{\sun})^{1/3}
[(1+z_\mathrm{vir})/11]^2$.
Thus, in the fiducial case
($f_\mathrm{disc}=0.5$ and $\epsilon_\star =0.3$),
the opacity is estimated as $\sim 2.9\times 10^2\mathcal{D}$.
Accordingly, if the dust-to-gas ratio is higher than $3\times 10^{-3}$,
the extinction significantly reduces the UV light. This explains the
inefficient radiation force for such a large dust-to-gas ratio.
For the photoelectric effect, photons at shorter wavelengths
(such as $\sim 0.1~\micron$) are important; the opacity is higher
by a factor of $\sim 5$ at such a short wavelength. This means
that even if $\mathcal{D}=10^{-3}$, the optical depth for
photons causing photoelectric emission becomes
of order unity. Thus, Coulomb drag is significantly suppressed if
$\mathcal{D}\gtrsim 10^{-3}$. From the above estimates,
we can confirm that $\mathcal{D}\sim 10^{-3}$ is the optimum
dust-to-gas ratio for grain escape (i.e.\ strong enough radiation
pressure and suppressed photoelectric charging).

\subsection{Effect of galaxy parameters}\label{subsec:gal_para}

\begin{figure}
\begin{center}
\includegraphics[width=0.45\textwidth]{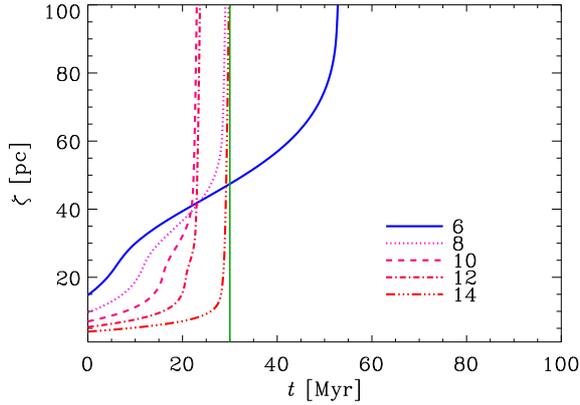}
\end{center}
\caption{Same as Fig.\ \ref{fig:vel_a}{d} but for various redshifts
{with $a=0.1~\micron$}.
The solid, dotted, dashed, dot--dashed, and triple-dot--dashed lines show the results
for $z_\mathrm{vir}=6$, 8, 10, 12, and 14, respectively.
\label{fig:vel_z}}
\end{figure}

Here we examine the parameters characterizing the galaxy.
First, we focus on the effect of formation redshift ($z_\mathrm{vir}$).
In Fig.\ \ref{fig:vel_z}, we show $\zeta$ as a function of $t$
for various $z_\mathrm{vir}$. We observe that the
dust successfully escapes within $t_\star =30$ Myr if $z_\mathrm{vir}\gtrsim 8$.
For galaxies with higher $z_\mathrm{vir}$, the gas is less
extended because of their lower scale height; moreover,
the grain velocity at low $\zeta$ is not sensitive to $z_\mathrm{vir}$
because both drag and radiation pressure scale with the
surface density (i.e.\ their scalings with $z_\mathrm{vir}$ are common).
Thus, grain escape occurs on a shorter time-scale for higher $z_\mathrm{vir}$.
%%The grain velocity is slower at low $\zeta$ for higher $z_\mathrm{vir}$ because
%%the gas is denser (i.e.\  drag is stronger; see equation \ref{eq:Sigmab}).
Note that the initial position is the exponential height, which depends on $z_\mathrm{vir}$
(as well as on $M_\mathrm{vir}$).
At high $\zeta$, higher-$z_\mathrm{vir}$ cases have higher grain velocities
because of higher stellar surface densities (i.e.\ higher radiation pressure)
for a fixed $M_\mathrm{vir}$.
For $z_\mathrm{vir}=6$, the grain velocity is kept low; thus, unless the star
formation activity lasts much longer than 30 Myr, grains do not
escape from the galactic disc. This means that high redshift such as $z\gtrsim 8$
is suitable for grain escape.

\begin{figure}
\begin{center}
\includegraphics[width=0.45\textwidth]{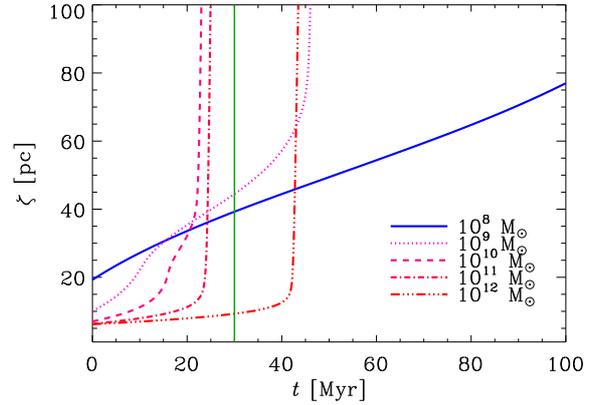}
\end{center}
\caption{Same as Fig.\ \ref{fig:vel_a}{d} but for various virial masses
{with $a=0.1~\micron$}.
The solid, dotted, dashed, dot--dashed, and triple-dot--dashed lines
show the results for $M_\mathrm{vir}=10^8$, $10^9$, $10^{10}$,
$10^{11}$ and $10^{12}~M_\odot$, respectively.
\label{fig:vel_M}}
\end{figure}

Next, we show the dependence on $M_\mathrm{vir}$ in Fig.\ \ref{fig:vel_M}.
There is a broad tendency that dust is more easily accelerated in
higher-$M_\mathrm{vir}$ objects in spite of their stronger gravity
(equation \ref{eq:beta}). Note that grain velocities at low $\zeta$ are
not sensitive to $M_\mathrm{vir}$ because both gas density (drag) and
stellar surface density (radiation pressure) are scaled in the same way
in terms of $M_\mathrm{vir}$. However, the gas extension toward the
$\zeta$ direction (i.e.\ scale height) is smaller in more massive galaxies. Therefore,
it is easier for dust grains to escape from dense regions in more massive
objects.
%%This is because the surface density of stars
%%is higher in higher-$M_\mathrm{vir}$ objects.
However, for $M_\mathrm{vir}=10^{12}$ M$_{\sun}$,
the grain escape takes more time: this is because dust extinction is higher
(UV is weaker) because of higher column density.
%%the gas density is also high, which leads to larger gas drag in the disc.
As estimated in Section \ref{subsec:dg}, the dust optical depth scales as
$\propto M_\mathrm{vir}^{1/3}$. This means that the dust opacity
for $M_\mathrm{vir}=10^{12}$ M$_{\sun}$ is
$10^{2/3}\simeq 4.6$ times higher than that for $M_\mathrm{vir}=10^{10}$ M$_{\sun}$.
Indeed, if we suppress the dust-to-gas ratio by a factor 4.6 ($\sim 0.002$), the dust grain
successfully escapes from the disc within $t_\star =30$ Myr.
However, it may not be easy to maintain such massive objects dust-poor (1/50
times the solar), since
they are located in high-density peaks in the Universe
(see also Section \ref{subsec:condition}). 
To summarize, with a fixed dust-to-gas ratio, a high mass does not necessarily
mean a more dust escape
from the disc. For low-mass ($M_\mathrm{vir}\lesssim 10^9$ M$_{\sun}$) objects,
grains are less efficiently accelerated because the gas is more extended
(i.e.\ for the same reason as in the case of low $z_\mathrm{vir}\sim 6$).
Therefore, massive galaxies with
$M_\mathrm{vir}\gtrsim 10^{10}$ M$_{\sun}$ are favourable for grain escape;
but extremely dust-poor ($\sim 1/50$ times the Milky Way dust-to-gas ratio)
condition is required for very
massive ($M_\mathrm{vir}\sim 10^{12}$ M$_{\sun}$) objects.

\begin{figure}
\begin{center}
\includegraphics[width=0.45\textwidth]{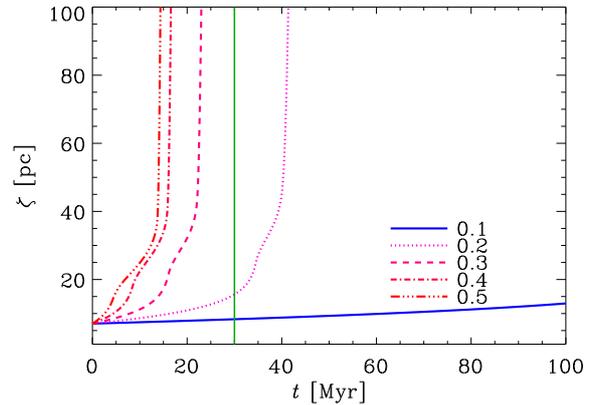}
\end{center}
\caption{Same as Fig.\ \ref{fig:vel_dg} but for various star formation
efficiencies {with $a=0.1~\micron$}.
The solid, dotted, dashed, dot--dashed, and triple-dot--dashed lines
show the results for $\epsilon_\star =0.1$, 0.2, 0.3, 0.4, and 0.5,
respectively.
\label{fig:vel_es}}
\end{figure}

We examine
the dependence on the star formation efficiency $\epsilon_\star$
in Fig.\ \ref{fig:vel_es}. As expected, grains are more efficiently
accelerated in the case of higher $\epsilon_\star$ because of higher radiation pressure.
In the case of $\epsilon_\star\leq 0.2$, the grain does not reach a high $\zeta\gtrsim 40$ pc
within $t_\star =30$ Myr.
Thus, a high star formation efficiency $\epsilon_\star\gtrsim 0.3$ (or
$\epsilon_\star f_\mathrm{disc}\gtrsim 0.15$ for the total conversion efficiency of
baryon into stars) is necessary for grain escape.

\begin{figure}
\begin{center}
\includegraphics[width=0.45\textwidth]{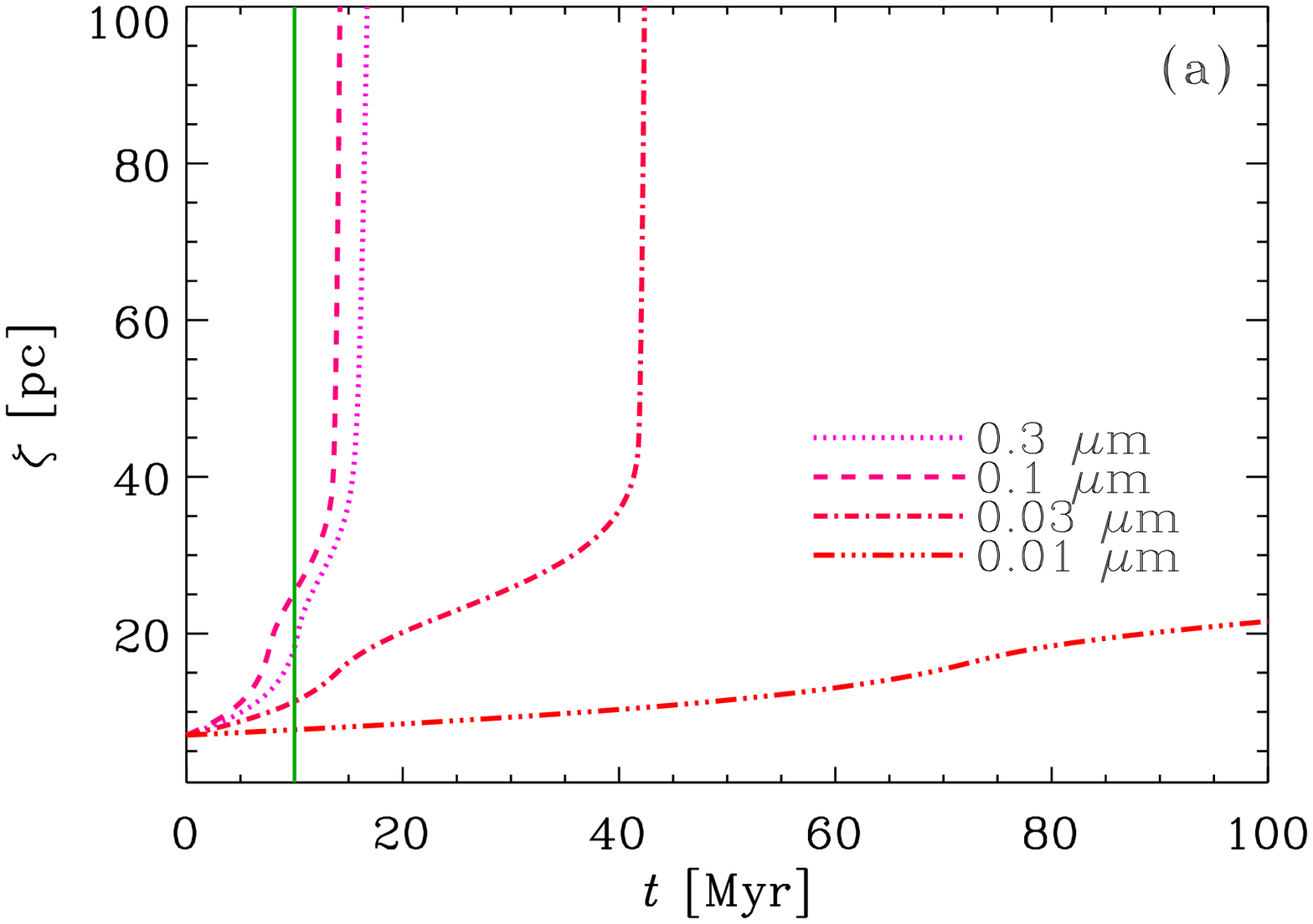}
\includegraphics[width=0.45\textwidth]{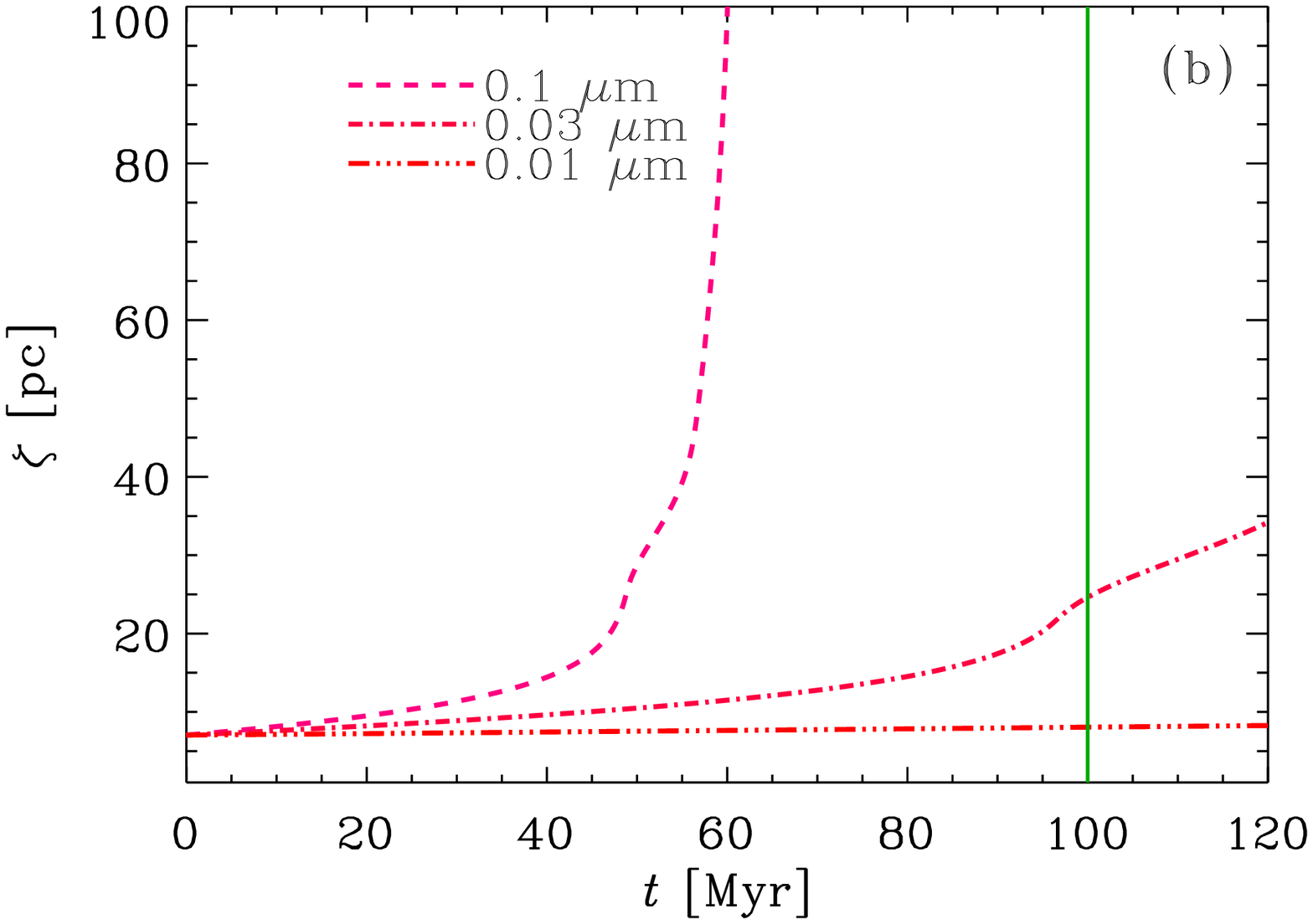}
\end{center}
\caption{Same as Fig.\ \ref{fig:vel_a}{d} but for different stellar ages,
$t_\star =10$ and 100~Myr for panels (a) and (b), respectively.
The different lines show different grain radii with the same line species
as in  Fig.\ \ref{fig:vel_a}{d} (also shown in the legend). Note that we do not show
$a=1~\micron$ ($a=1$ and 0.3 $\micron$) in panel (a) [panel(b)]
because the grain velocity becomes negative.
\label{fig:vel_age}}
\end{figure}

In Fig.\ \ref{fig:vel_age}, we show the $\zeta$--$t$
relation for different stellar ages $t_\star$ and different grain radii.
We observe that the outward motion tends to be slower for
older stellar ages.
Note that we fixed the total stellar mass formed (thus,
a larger $t_\star$ means lower star formation rate, i.e.\ lower UV luminosity).
Grains with
$a\sim 0.1~\micron$ are pushed efficiently by radiation pressure
and escape from the disc within $t_\star$ for $t_\star =100$ Myr.
In contrast, in the case of $t_\star =10$ Myr, the grain is
difficult to escape within the short age.
Thus, we conclude that sustaining star formation activity for a
long ($\gtrsim 30$ Myr) time is necessary to push the dust grains
continuously out of the galactic disc.

\section{Discussion}\label{sec:discussion}

\subsection{Condition for grain escape}\label{subsec:condition}

We have shown that only grains with $a\sim 0.1~\micron$ can be
transported to the halo. Smaller ($a\lesssim 0.03~\micron$) and
larger ($a\gtrsim 1~\micron$) grains are efficiently trapped in the
galactic disc or at disc--halo interface because of gas drag and gravity,
respectively. Graphite is more efficiently pushed by
radiation pressure at low $\zeta$ than silicate, while it is less efficiently
accelerated in the halo. We have also found that
there is an optimum dust-to-gas ratio, $\mathcal{D}\sim 10^{-3}$, for
grain escape. Galaxies with $z_\mathrm{vir}\gtrsim 8$ are efficient in
pushing dust out of the disc.
There is also an optimum virial mass range for grain escape,
$M_\mathrm{vir}\sim 10^{10}$--$10^{11}$ M$_{\sun}$.
In order to achieve a sufficient stellar brightness for grain escape,
more than 15 per cent of the baryon should be converted to the
stars in the current star formation episode, and the star formation should
continuously last longer than $\sim 30$ Myr. These conditions can be used
to model the dust supply to galaxy halos at high redshift in an extended
framework such as a semi-analytic model.

%%Although the results are sensitive to the galaxy parameters in terms of
%%the $\zeta$ dependence, the velocities at small and large $\zeta$ do not
%%strongly depend on $z_\mathrm{vir}$ and $M_\mathrm{vir}$.
%%Indeed, as we observe in Figs.\ \ref{fig:vel_z} and \ref{fig:vel_M},
%%$v_\zeta\sim 0.1$--0.3 km s$^{-1}$ at small $\zeta$ while
%%$v_\zeta$ approaches $>10$ km s$^{-1}$ at large $\zeta$.
%%Thus, we do not observe very large difference in the grain velocity
%%among various values of $z_\mathrm{vir}$ and $M_\mathrm{vir}$.
%%This is due to some compensating effects.
%%When the stellar density is high, the gas density is also high
%%(with a fixed $\epsilon_\star$); that is, radiation pressure and
%%gas drag both become high. Therefore, the velocity at low $\zeta$
%%determined by the balance between gas drag and radiation pressure
%%is not sensitive to the parameters. At large $\zeta$, the velocity approaches
%%the value determined by the balance between gravity and radiation pressure,
%%both of which are scaled with the surface density. Because of this common
%%scaling between gravity and radiation pressure, the grain velocity at
%%large $\zeta$ is not sensitive to $z_\mathrm{vir}$ and $M_\mathrm{vir}$.

We have found that the optimum dust-to-gas ratio is determined by the
optical depth at UV wavelengths. If the dust-to-gas ratio is too high,
the stellar UV light is heavily extinguished so that radiation pressure
becomes too weak to push the grain. The condition for the
weak extinction is determined by the optical depth $<1$ at a typical UV
wavelength for stellar radiation ($\sim 0.2~\micron$). The optical
depth at 0.2 $\micron$ is estimated as
$\tau_{0.2}\sim 8.3\times 10^{2}\mathcal{D}(1-\epsilon_\star )f_\mathrm{disc}
(M_\mathrm{vir}/10^{10}~\mathrm{M}_{\sun})^{1/3}
[(1+z_\mathrm{vir})/11]^2$ (Section \ref{subsec:dg}).
If this is smaller than 1, the extinction effect is not significant.
If the dust-to-gas ratio is too low, stellar UV emission is not attenuated,
so that it efficiently charges grain positively. This enhances
Coulomb drag, and suppresses the grain velocities. Thus, it is
desirable that the optical depth at a
typical wavelength of the photoelectric effect ($\sim 0.1~\micron$) is
larger than 1 (corresponding optical depth at $\sim 0.2~\micron$
is roughly 1/5). 
Therefore,
the optimum dust-to-gas ratio, denoted as $\mathcal{D}_\mathrm{optimum}$,
is determined by the condition $0.2<\tau_{0.2}<1$. This is translated into
the following condition:
\begin{align}
\mathcal{D}_\mathrm{optimum} &\sim (0.6\mbox{--}3)\times 10^{-3}
\left(\frac{1-\epsilon_\star}{0.7}\right)^{-1}\left(\frac{f_\mathrm{disc}}{0.5}\right)^{-1}
\left(\frac{M_\mathrm{vir}}{10^{10}~\mathrm{M}_{\sun}}\right)^{-1/3}\nonumber\\
&\times \left(\frac{1+z_\mathrm{vir}}{11}\right)^{-2}.
\end{align}
Thus, the optimum dust-to-gas ratio is roughly $\mathcal{D}_\mathrm{eff}\sim 10^{-3}$
in the fiducial condition, but the above formula can be used for other
parameter values.

For $M_\mathrm{vir}=10^{10}$--$10^{11}$ M$_{\sun}$,
the corresponding stellar mass with $f_\mathrm{disc}=0.5$ and
$\epsilon_\star =0.3$ is $2\times 10^8$--$2\times 10^9$ M$_{\sun}$.
According to a semi-analytic model by \citet{Popping:2017aa},
the typical dust-to-gas ratio in the above stellar mass range is
$\sim 3\times 10^{-5}$--$3\times 10^{-3}$ at $z\gtrsim 7$, overlapping with the
above optimum dust-to-gas ratio. However, for
$M_\mathrm{vir}=10^{12}$ M$_{\sun}$, the optimum dust-to-gas ratio
is $\lesssim 10^{-4}$, but it is difficult to maintain the dust-to-gas ratio of such a massive
galaxy lower than $10^{-4}$ \citep{Popping:2017aa}.
The above dust-to-gas ratio would indicate a metallicity of 1/100 Z$_{\sun}$
if we assume a proportionality between dust-to-gas ratio and metallicity,
and a theoretical stellar-mass--metallicity relation at $z\sim 10$
indicates that a galaxy with the above stellar mass range has a higher
metallicity \citep{Torrey:2019aa}.

\subsection{Dynamical effects of radiation pressure}
\label{subsec:effect}

%%First, we showed that the effect of radiation pressure on
%%the hydrostatic structure of the gas disc is negligible.
%%This does not deny the importance of local intense radiation
%%field arising from H \textsc{ii} regions on local kinematics within
%%the disc (Ferrara 1991; Ishiki et al.). However, it is unlikely
%%that the radiation field drastically change the structure of the
%%gaseous disc. Supernova feedback would rather be important for
%%significantly modifying the gas structures.

The grain motion have different characteristic behaviours
at different heights from the disc plane.
At low $\zeta\ll H_\mathrm{g0}$, grains have typically
$v_\zeta\sim 0.1$ km s$^{-1}$
%%, and lie in the `escaping' region
%%(i.e.\  out of the shaded regions in Fig.\ \ref{fig:vel_a}a)
unless the grain radius is as large as $\sim 1~\micron$ (Fig.\ \ref{fig:vel_a}a).
This means that the radiation pressure causes dust--gas decoupling on a
vertical scale of
$\sim v_\zeta t_\star\sim 3.1(v_\zeta /0.1~\mathrm{km~s}^{-1})
(t_\star /30~\mathrm{Myr})$ pc. Thus, it is important to note that radiation
pressure produces a `drift' of dust grains relative to the
gas on a spatial scale of $\sim$ a few pc. The implication of this is
that, if we calculate grain dynamics on scale $\lesssim$ a few pc on
a time-scale of a few tens Myr ($\sim$ a typical duration of a star formation
episode),
we need to take into account the effect of radiation pressure on
the grain motion.

At intermediate $\zeta\sim\mbox{a few}\times H_\mathrm{g0}$,
although the density drops significantly,
grain motions are
strongly affected by gas drag because of the grain charge.
Therefore, even if the gas density
drops, the grain velocities stay around 1 km s$^{-1}$.
Larger grains have larger velocities because of their
lower surface-to-volume ratio (less drag per grain mass)
and larger $Q_\mathrm{rad}$. However, if the grains are too large,
gravity suppresses the grain velocity. Therefore, there is an optimum
grain radius $\sim 0.1~\micron$ for grain escape from the disc.

At large $\zeta\gtrsim 10H_\mathrm{g0}$, grains with $a\sim 0.1~\micron$
are rapidly accelerated to
$>10$ km s$^{-1}$. At this stage, gas drag can become so weak that
grains are accelerated freely by radiation pressure.
Note that, since both gravity and radiation pressure are constant at large $\zeta$
{in our approximation of an infinitely wide disc}, the grains
continue to be accelerated outward once they escape from the disc.
Thus, if a grain with $a\lesssim 0.1~\micron$ somehow reaches
$\zeta\sim 10H_\mathrm{g0}$, it is successfully injected into the halo.

\subsection{Observational implications for the grain size in galaxy halos}

\citet{Hirashita:2018aa} showed that galaxy halos at $z\lesssim 2$
contain dust grains with
$a\sim 0.01$--0.03 $\mu$m, based on the reddening curves of objects tracing
galaxy halos (such as Mg \textsc{ii} absorbers; \citealt{Menard:2012aa}).
According to the results above, the grains with $a\sim 0.01~\mu$m
do not escape from the galaxy disc, while larger grains with $a\sim 0.1~\mu$m
can be supplied to the halo. Therefore, the dust transport mechanism by
radiation pressure has
difficulty in explaining the small grains derived by \citet{Hirashita:2018aa}.
We should note that
hydrodynamical motion driven by SN feedback (i.e.\ galactic wind)
could also transport dust grains to halos.
\citet{Hou:2017aa}, using a simulation of single disc galaxy,
showed that the grains transported by stellar feedback
are biased to large ($>0.03~\mu$m) sizes. This is because dust grains
formed by SNe have large sizes and they are transported before
they are processed in the ISM by shattering. \citet{Aoyama:2018aa}
confirmed this conclusion by a cosmological simulation.
Therefore, SN feedback does not seem to provide small dust grains
to galaxy halos.

Dust may be processed in the CGM or IGM.
It may be destroyed by sputtering, which, however, tends to
destroy small grains more efficiently than large ones
\citep{Draine:1979aa,Nozawa:2006aa}.
%%Moreover, sputtering conserves the number of grains as long as it only make the grains
%%smaller.
Thus, sputtering generally has
difficulty in producing a grain size distribution in excess of small
grains. Indeed, \citet{Bianchi:2005aa} showed that the grain size
distribution after sputtering in the IGM is still flat with a shift to
smaller radii. Such a grain size distribution cannot explain
the strongly rising trend of the reddening in Mg \textsc{ii}
absorbers toward short UV wavelengths.

The  observational evidence of small grains in the regions out of galaxies
is based on Mg \textsc{ii} absorbers and the CGM,
which could have some density contrast.
\citet{Lan:2017aa} proposed that Mg \textsc{ii} absorbers are associated with
clouds with a gas density of $\sim 0.3$ cm$^{-3}$. This density is high enough
for grain--grain collisions to occur on a time-scale
of $\sim 1$ Gyr \citep{Aoyama:2017aa}, which is shorter than
the interval between $z\sim 10$, in which our model is interested, and
$z\sim 2$, where the Mg \textsc{ii}
absorbers are sampled. Therefore, the small grains
may have been formed by shattering in the circum-galactic environment.

Eventually, we have to combine the above discussions with the
evolution of grain size distribution in a galactic disc.
The grain size distribution in the early stage of galaxy evolution
is biased to large ($a\gtrsim 0.1~\micron$) sizes because
dust grains produced by stellar sources are considered to be large
\citep[e.g.][]{Nozawa:2007aa,Bianchi:2007aa,Yasuda:2012aa,DellAgli:2017aa}.
Therefore, we expect that grains with $a\sim 0.1~\micron$ exist
even in the early phase of galaxy evolution.
On the other hand, we have shown in Section \ref{subsec:dg} that
the optimum dust-to-gas ratio for grain escape is $\mathcal{D}\sim 10^{-3}$,
which is much less than the
dust-to-gas ratio in solar-metallicity environments ($\mathcal{D}\sim 10^{-2}$).
If we assume a rough proportionality between dust-to-gas ratio and metallicity ($Z$),
grain escape occurs most efficiently at $Z\sim 0.1$ Z$_{\sun}$.
It is interesting to note that this metallicity also corresponds to the
metallicity level at which the grain size distribution is strongly modified by
the interstellar processing \citep[e.g.][]{Asano:2013aa}.
Therefore, it is desirable in the future to solve the combined effects between
the evolution of grain size distribution in the galactic disc and the dust transport
to the halo.

\subsection{Implication for observed galaxy populations at high
redshift}

In Section \ref{subsec:gal_para}, we have shown that the star formation
efficiency $\epsilon_\star$ higher than 0.3 (or the baryonic mass fraction
converted to stars $\epsilon_\star f_\mathrm{disc}\gtrsim 0.15$) is required
to efficiently push the dust by radiation pressure. In the previous subsection,
we have also argued that star formation needs to last $\gtrsim 30$ Myr
to transport the dust to the halo. These numbers enable us to estimate
the SFR and stellar mass in galaxies whose halos are enriched with dust.

Our formulation is based on disc geometry, so that the
surface density was useful. Here, we need global quantities.
The total stellar mass, $M_\star =(\pi r_\mathrm{disc}^2)\Sigma_\mathrm{b}$,
is estimated as (see equations \ref{eq:Sigma_b} and \ref{eq:Sigma_star})
\begin{align}
M_\star &= \epsilon_\star (\Omega_\mathrm{b}/\Omega_\mathrm{M})f_\mathrm{disc}
M_\mathrm{vir}\nonumber\\
&= 2.0\times 10^8\left(\frac{\epsilon_\star}{0.3}\right)
\left(\frac{f_\mathrm{disc}}{0.5}\right)
\left(\frac{M_\mathrm{vir}}{10^{10}~M_\odot}\right)~M_\odot .\label{eq:Mstar_obs}
%%2.000
\end{align}
The star formation should last $\gtrsim 30$~Myr for grain escape. This, combined
with the above stellar mass, indicates SFR
$\lesssim 7(M_\mathrm{vir}/10^{10}~\mathrm{M}_{\sun})$ M$_{\sun}$ yr$^{-1}$.
LBGs typically have a comparable or larger mass than the
above \citep[e.g.][]{Bouwens:2016aa}.
\citet{Hashimoto:2018aa} analyzed the SED of a galaxy at $z=9.1$,
and found that it experienced an old star formation episode, which
lasted $\sim$100 Myr at $z\sim 15$--12. The established stellar
mass by this star formation activity is $\sim 10^9$ M$_{\sun}$, which is larger
than the value estimated in equation (\ref{eq:Mstar_obs}).
Thus, it is expected that this object had sufficiently strong radiation pressure
for halo enrichment with dust (with grain radius $\sim 0.1~\micron$).
If such a galaxy is prevalent at $z\sim 10$, we should consider
dust enrichment in galaxy halos by radiation pressure.
We need a statistical sample of galaxies at $z\sim 10$ to draw a
definite conclusion.

In our one-dimensional framework, it is difficult to predict how much
dust is supplied to halos. Nevertheless, the following rough estimate
is possible.
We consider a halo with $M_\mathrm{vir}=10^{10}~M_\odot$.
With $f_\mathrm{disc}=0.5$, the gas mass in the halo is estimated as
$6.7\times 10^8$ M$_{\sun}$, and that in the disc as the same amount.
We have shown that dust grains with
$a\sim 0.1~\micron$ escape from the galactic disc.
Grains with $a\sim 0.3~\micron$ could marginally escape
(Figs.\ \ref{fig:vel_a}{d} and \ref{fig:vel_gra_a}).
Adopting
an MRN grain size distribution ($\propto a^{-3.5}$) with lower and upper grain radii being
$0.005~\mu$m and 0.25 $\mu$m, respectively, we estimate that
the mass fraction of dust with $a\geq 0.1~\micron$ is 0.43.
%%0.4280
%%If the grain size distribution is biased to large sizes as expected for
%%the early phase of galaxy evolution (Section \ref{subsec:effect}), the fraction would become
%%even larger.
As shown above, the dust-to-gas ratio at which the radiation pressure works
the most efficiently is $\mathcal{D}\sim 10^{-3}$ (Section \ref{subsec:dg}).
If we assume a star formation efficiency of 0.3, the
remaining gas mass in the disc is $4.7\times 10^8$~M$_{\sun}$, which
indicates that the total dust mass is $\sim 4.7\times 10^5$ M$_{\sun}$ with
$\mathcal{D}=10^{-3}$.
Therefore, the dust mass transported
to the halo is expected to be $\sim 2.0\times 10^5$ M$_{\sun}$.
%%Since the total gas mass in the halo is
%%$\sim 5\times 10^8~M_\odot$, the dust-to-gas ratio in the halo is
%%$\sim 2\times 10^{-4}$.
The total stellar mass, on the other hand, is
$2.0\times 10^8$ M$_{\sun}$. Thus,
the mass ratio of the halo dust to the stars is
$\sim 10^{-3}$. From an observational point of view, \citet{Hirashita:2018aa},
based on \citet{Menard:2010aa}, argued that the observed mass ratio of
the halo dust to the stars is $\sim 10^{-3}$ for low-redshift galaxies.
Comparing the numbers, high-redshift halos could be as dust-rich as low-redshift halos.
Therefore, dust transport by radiation pressure
in high-redshift galaxies
should be considered to understand the origin of the dust in halos.

If grains with $a\sim 0.1~\micron$ efficiently escape out of the galactic disc,
this could have a significant influence on the dust evolution in the galactic disc.
First, the dust abundance in the galactic disc could be decreased or the dust distribution
is extended toward the galaxy halo. Such an extended dust component will have
a lower temperature than that in the disc, because it is farther from the
radiation sources (i.e.\ stars in the disc).
%%Interferometers like ALMA are not capable of detecting extended dust emission.
Since lower-temperature dust emits less radiation, extended dust distribution
could be a reason for
non-detection of dust in a large fraction of high-redshift ($z\gtrsim 7$) galaxies by ALMA.
Second, the remaining dust in the galactic disc could be biased to
small ($a\lesssim 0.03~\micron$) grains because large ($a\sim 0.1~\micron$)
grains are preferentially transported out of the disc. This could produce a steep extinction curve.
Indeed, as shown by \citet{Hashimoto:2018ab}, those galaxies at $z>6.5$ whose
dust emission is not detected by ALMA tend to have
steeper extinction curves, if their positions in the so-called IRX--$\beta$ diagram
are interpreted
as reflecting the extinction (attenuation) curve. Selective loss of relatively large
($a\sim 0.1~\micron$) grains could explain this tendency.

\subsection{Uncertainties}\label{subsec:uncertainties}

We simplified the problem to make it analytically tractable.
It is worth mentioning the limitations and uncertainties caused by
some simplifications.

The initial velocity and position are not important as long as
it is set at $\zeta\lesssim 20$~pc (several scale heights), since the grain velocity is
determined by the equilibrium value mainly determined by the
balance between gas drag and radiation pressure
(Section~\ref{subsec:grain_param}; Fig.~\ref{fig:vel_a}a).
For example, the typical drag time-scale in the galactic disc is
$\sim 10^3$ yr; thus, even if the initial grain velocity is 1000 km s$^{-1}$,
the grain only moves 1 pc. This means that the grains are efficiently
decelerated by gas drag on a scale much smaller than the
disc scale height.
Therefore, our results are robust against the initial velocity.
In contrast, the initial velocity is important if the initial position is above
several scale heights.

The vertical gas structure is determined by the gravity which is
characterized by $\beta$ and $H_\mathrm{M}$. These parameters are scaled with
$M_\mathrm{vir}$ and $z_\mathrm{vir}$ (equations \ref{eq:beta} and \ref{eq:H_M}).
These scalings and the associated normalizing factors might be too simple.
However, the relative strength between gas drag and radiation pressure
is rather robust because both scale with the surface densities of the galactic disc.
As long as we fix the star formation efficiency, thus,
the relative strength between gas drag and radiation pressure, which broadly
determines the condition for grain escape, is not sensitive to the assumed scalings.
For the same reason, the condition for grain escape is not sensitive to the
change of $q_\mathrm{disc}$.
%%Indeed, as shown in Fig.\ \ref{fig:vel_z_norm}, grain velocities are not very sensitive
%%if $\zeta$ is normalized to the typical scale height $H_\mathrm{g0}$.

The scale height is also governed by the velocity dispersion of the gas,
$\sigma$. The change of $\sigma$ does not
affect our conclusions. If we assume a smaller $\sigma$, the gas density
becomes $\sigma^2$ times larger, so that the velocity is suppressed
roughly by a factor of $\sigma^2$. However, the disc is also $1/\sigma^2$ times
thinner. Therefore, the time-scale of grain escape from the disc is not sensitive
to $\sigma$.

If the grains are charged and magnetic field is present,
the grain motion is also affected by the
Lorentz force. However, it is not clear yet whether magnetic fields are
amplified efficiently in high-redshift galaxies. The role of magnetic field is
to constrain the motion of charged grains to the direction of magnetic field. Therefore,
as long as the magnetic field has a significant poloidal component (or
a component open to the $\zeta$ direction), the grain escape condition
is not significantly affected so that the
results obtained in this paper is still applicable for the grain motion
in the $\zeta$ direction.

As mentioned at the beginning of Section \ref{sec:result},
our simple model is not capable of treating inhomogeneity
of gas structure in terms of ionization degree and temperature.
The metallicity (and dust-to-gas ratio) could also be strongly inhomogeneous
\citep{Pallottini:2017aa}.
In the future work, it is desirable to solve these features, but it should
also be kept in mind that, to this goal, we need to make a comprehensive
model of hydrodynamics (including star formation and stellar feedback)
and radiation transfer. Our simple models in this paper gives a first guide
to such more complicated models, or provides a basis on which we
model dust supply from galactic discs to halos in semi-analytic models.

\section{Conclusion}\label{sec:conclusion}

For the purpose of clarifying the origin of dust in galaxy halos or in the CGM/IGM,
we have investigated dust transport from the galactic disc to the halo
(`grain escape') by
radiation pressure in high-redshift galaxies.
We have considered radiation pressure
arising from the star formation activities and focused on the redshift range
where recent observations indicate
the occurrence of active star formation ($z\sim 10$).
We have solved the motion of a grain
with various sizes considering radiation pressure, gas drag, and gravity
in the vertical direction of the galactic disc. Radiation pressure is estimated
in a consistent manner with the stellar SED and dust extinction.

We give the virial mass $M_\mathrm{vir}$ and the formation redshift
$z_\mathrm{vir}$ as parameters. First, we focus on
$M_\mathrm{vir}=10^{10}$ M$_{\sun}$ and $z_\mathrm{vir}=10$.
We point out that grain charging by UV light plays an important role in
gas drag at a few--10 scale heights of the galactic disc and that the
grain velocities are suppressed to $\sim 1$ km s$^{-1}$.
Graphite grains are slightly easier to escape from the disc than
silicates because they receive more radiation force.
However, graphite has slightly lower velocity in the halo than silicate
because of its larger charge (stronger drag).
We find that grains with radius $a\sim 0.1~\micron$
successfully escape from the galactic disc if the current star formation
episode converts more than 15 per cent of
the baryon content into stars and lasts $\gtrsim 30$ Myr.
Larger ($a\gtrsim 1~\micron$) grains are efficiently trapped in the disc
because of its large inertia (gravity), while small ($a\lesssim 0.01~\micron$)
grains are also strongly influenced by gas drag because of their low
efficiency of receiving radiation force.

Next we vary $M_\mathrm{vir}$ and $z_\mathrm{vir}$ and examine if a
grain with $a=0.1~\micron$ could escape from the disc.
High-redshift galaxies ($z_\mathrm{vir}\gtrsim 8$)
are favourable for grain escape because of their
lower scale height (i.e.\ it takes less time to cross the dense region).
For the same reason, more massive galaxies are favorable for grain escape;
however, we also find that if the galaxy is too massive ($M_\mathrm{vir}\gtrsim 10^{12}$ M$_{\sun}$),
dust extinction is high enough to extinguish a significant fraction of stellar UV
with $\mathcal{D}\sim 10^{-3}$. Maintaining a condition with
$\mathcal{D}\ll 10^{-3}$ (or $Z\ll 0.1$ Z$_{\sun}$) in such a massive
object may be difficult.
Thus, we argue that there is an optimum range of the virial mass,
$M_\mathrm{vir}\sim 10^{10}$--$10^{11}$ M$_{\sun}$,
for grain escape.
We estimate that the dust mass in the halos of these galaxies could
reach $10^{-3}$ times the stellar mass in the disc, which is comparable to
the dust abundance found in the CGM at $z\lesssim 3$.
A recently found galaxy at $z=9.1$ {\citep{Hashimoto:2018aa}}
satisfies the condition of grain escape
for $a\sim 0.1~\micron$, implying that
dust injection to halo by radiation pressure already occurred at $z\sim 10$.
Therefore, we conclude that radiation pressure in high-$z$ galaxies is important
in considering the origin of dust in galaxy halos.

\section*{Acknowledgements}

We are grateful to W.-H. Shao, A. Ferrara, and M. Murga for useful discussions
and to the anonymous referee for helpful comments.
HH thanks the Ministry of Science and Technology for support through grant
MOST 105-2112-M-001-027-MY3 and MOST 107-2923-M-001-003-MY3
(RFBR 18-52-52-006).
This work was supported by NAOJ ALMA Scientific Research Grant Number 2016-01A.



%\bibliographystyle{mnras}
\bibliographystyle{mnras}
\bibliography{/Users/hirashita/bibdata/hirashita}
%%\bibliography{hirashita}

%%%%%%%%%%%%%%%%% APPENDICES %%%%%%%%%%%%%%%%%%%%%

\appendix

\section{Relevant quantities}\label{app:age}

For convenience, we give numerical estimates of some quantities
used in the text.

The virial radius, $r_\mathrm{vir}$, is numerically estimated based on
equation (\ref{eq:rvir}) as
\begin{align}
r_\mathrm{vir}=6.0\left(\frac{M_\mathrm{vir}}{10^{10}~\mathrm{M}_{\sun}}\right)^{1/3}
\left(\frac{1+z_\mathrm{vir}}{11}\right)^{-1}~\mathrm{kpc}.\label{eq:rvir_num}
%%6.003
\end{align}
The disc radius is accordingly
\begin{align}
r_\mathrm{disc}=1.1\left(\frac{q_\mathrm{disc}}{0.18}\right)
\left(\frac{M_\mathrm{vir}}{10^{10}~\mathrm{M}_{\sun}}\right)^{1/3}
\left(\frac{1+z_\mathrm{vir}}{11}\right)^{-1}~\mathrm{kpc}.\label{eq:rdisc}
%%1.080
\end{align}
The virial temperature ($T_\mathrm{vir}$) is estimated as
\begin{align}
T_\mathrm{vir} &\simeq
\frac{G\mu' m_\mathrm{H}M_\mathrm{vir}}{3k_\mathrm{B}r_\mathrm{vir}}
\nonumber\\
&= 1.9\times 10^5\left(\frac{M_\mathrm{vir}}{10^{10}~\mathrm{M}_{\sun}}\right)^{2/3}
\left(\frac{1+z_\mathrm{vir}}{11}\right)~\mathrm{K},\label{eq:Tvir}
%%1.880
\end{align}
where $\mu ' m_\mathrm{H}$ is the mean particle mass ($\mu' =0.65$),
and we used equation (\ref{eq:rvir}) from the first to the second line.

The cosmic age at high redshift is
approximated by \citep[e.g.][]{Furlanetto:2006aa}
\begin{align}
t_0(z)\simeq \int_\infty^z\frac{dz'}{(1+z')H(z')}=
4.7\times 10^8\left(\frac{1+z}{11}\right)^{-3/2}~\mathrm{yr},\label{eq:cosmic_age}
%%4.659
\end{align}
where $H(z)$ is the Hubble parameter at redshift $z$, which
is approximated in the matter-dominated Universe at high redshift as
$H(z)\simeq H_0\Omega_\mathrm{m}^{1/2}(1+z)^{3/2}$.

% Don't change these lines
\bsp	% typesetting comment
\label{lastpage}
\end{document}